\documentclass[twocolumn]{aastex63}
\usepackage{graphicx}
\usepackage{multirow,tabularx}
\usepackage{chngcntr, array}

\newcommand{\romannumeralcap}[1]
    {\MakeUppercase{\romannumeral #1}}
\newcommand{\hstcos}{\textit{HST}/COS}
\newcommand{\emcee}{\textsc{emcee}}

\newcolumntype{P}[1]{>{\centering\arraybackslash\hspace{0pt}}p{#1}}

\received{5/16/23}
\accepted{9/28/23}

\submitjournal{ApJ}

\shorttitle{SESAMME}
\shortauthors{Jones et al.}
\graphicspath{{./}{figures/}}

\begin{document}

\title{Simultaneous Estimates of Star-cluster Age, Metallicity, Mass, and Extinction (SESAMME) \romannumeralcap{1}: Presenting an MCMC Approach to Spectral Stellar Population Fitting}

\correspondingauthor{Logan Jones}
\email{lojones@stsci.edu}

\author[0000-0002-1706-7370]{Logan H. Jones}
\affiliation{Space Telescope Science Institute, 3700 San Martin Drive, Baltimore, MD 21218, USA}

\author[0000-0003-4857-8699]{Svea Hernandez}
\affiliation{AURA for ESA, Space Telescope Science Institute, 3700 San Martin Drive, Baltimore, MD 21218, USA}

\author[0000-0002-0806-168X]{Linda J. Smith}
\affiliation{Space Telescope Science Institute, 3700 San Martin Drive, Baltimore, MD 21218, USA}

\author[0000-0003-4372-2006]{Bethan L. James}
\affiliation{AURA for ESA, Space Telescope Science Institute, 3700 San Martin Drive, Baltimore, MD 21218, USA}

\author[0000-0003-4137-882X]{Alessandra Aloisi}
\affiliation{Space Telescope Science Institute, 3700 San Martin Drive, Baltimore, MD 21218, USA}

\author[0000-0003-0069-1203]{S\o{}ren Larsen}
\affiliation{Department of Astrophysics/IMAPP, Radboud University, PO Box 9010, 6500 GL Nijmegen, The Netherlands}

\begin{abstract}

We present the first version release of SESAMME, a public, Python-based full spectrum fitting tool for Simultaneous Estimates of Star-cluster Age, Metallicity, Mass, and Extinction. SESAMME compares an input spectrum of a star cluster to a grid of 
stellar population models with an added nebular continuum component, using Markov chain Monte Carlo (MCMC) methods to sample the posterior probability distribution in four dimensions: cluster age, stellar metallicity $Z$, reddening $E(B-V)$, and a normalization parameter equivalent to a cluster mass. SESAMME is highly flexible in the stellar population models that it can use to model a spectrum; our testing and initial science applications use both BPASS and Starburst99.
We illustrate the ability of SESAMME to recover accurate ages and metallicities even at a moderate signal-to-noise ratio (S/N $\sim 3 - 5$ per wavelength bin) using synthetic, noise-added model spectra of young star clusters. Finally, we test the consistency of SESAMME with other age and metallicity estimates from the literature using a sample of {\em HST}/COS far-UV spectra towards young, massive clusters in M83 and NGC 1313. We find that, on the whole, SESAMME infers star cluster properties that are consistent with the literature in both low- and high-metallicity environments.

\end{abstract}


\section{Introduction}
\label{sec:intro}

An accurate understanding of the current and past lives of galaxies requires a complete census of their elemental abundances. Innumerable works have examined the global and spatially-resolved metallicities of star-forming galaxies (SFGs), as well as metallicity gradients, usually through the analysis of emission lines from H{\sc~ii} regions \citep[e.g.,][and references therein]{Maiolino19}. These studies have also led to the development of global relations between a galaxy's metal content and other physical characteristics \citep[e.g., the mass-metallicity relation or MZR;][]{Zaritsky94,Tremonti04}. Taken together, these relations and metallicity gradient estimates are frequently used to investigate a galaxy's star-formation and merger history, the impact of galactic winds, and the flow of pristine gas from the intergalactic medium onto galaxies \citep[e.g.,][]{Ho15,Kudritzki15}. 

However, relying solely on the emission from ionized gas to characterize the chemical composition of SFGs may result in an incomplete picture of their metal content. The nebular emission from H{\sc~ii} regions, for example, can only be sustained for the $\sim$10 Myr period in which the most massive and UV-bright stars can photoionize their surrounding gas. As such, the metallicity of this gas is associated more with the most recent episodes of star formation and may be more enriched than the surrounding interstellar medium \citep[ISM;][]{Kunth86,Lebouteiller13}. Alternative perspectives on the metal content of SFGs can be obtained through analyses of individual stars or star clusters, which retain some memory of the chemical composition of their birth cloud. 
For very nearby galaxies ($D \sim$a few Mpc), measurements of stellar metallicities have been achieved using spectroscopic observations of blue and red supergiants \citep[e.g.,][]{Davies10,Davies15,Davies17,Kudritzki12,Kudritzki14,Gazak14,Hosek14,Bresolin16}, while other works use integrated-light (IL) spectroscopy of star clusters and stellar populations \citep[e.g.,][]{Halliday08,Mcwilliam08,Colucci11,Colucci12,Larsen12,Steidel16,chisholm19}. 

A spectral fitting technique developed by \citet[hereafter L12]{Larsen12} has been widely used in the literature to model the integrated light (IL) of star clusters \citep{Larsen12,Larsen18b,Hernandez17,Hernandez19,Hernandez21,Hernandez22,
Asad22,Gvozdenko22}.  In brief, the method involves first deriving stellar parameters of individual stars via color-magnitude diagrams or, more typically, theoretical isochrones of a given age and metallicity, from which one can synthesize stellar atmosphere models and mock spectra for each star. The sum of these synthetic spectra creates an IL spectrum of the cluster, which is then compared to the data. This process is iterated until an optimal fit is obtained for the individual elemental abundances across predefined wavelength bins. The L12 technique was originally developed to study the properties of extragalactic star clusters with very high spectral resolution ($R \sim$30,000) optical observations, but has since been successfully applied to spectra at lower resolutions $R \sim 8000$ \citep{Hernandez17} and at both infrared \citep{Larsen18b} and UV wavelengths \citep{Hernandez19}.

However, \citet{Hernandez19} report a caveat to the L12 process that becomes especially prevalent when dealing with UV observations. When applying the L12 method to {\em HST}/COS FUV spectra of a sample of young massive clusters (YMCs) in M83, they find that varying the assumed age of the input isochrone can cause significant changes in the inferred stellar metallicities. For clusters with relatively uncertain ages from, e.g., spectral energy distribution  (SED) fitting, dialing the assumed age up or down by the $1\sigma$ uncertainties induced changes in their inferred metallicities of order $\sim$0.2 -- 0.3 dex on average and as high as 0.4 dex in individual clusters. Additionally, the scatter of metallicity measurements from different wavelength bins was consistently higher when assuming the $-1\sigma$ age. They attribute these effects to the rapid evolution in the UV properties of massive star clusters in the first 10 Myr of their lives, reinforcing the need for precise age estimates for very young star clusters. 

In this work, we introduce a complement to the L12 method in the form of a Python tool for Simultaneous Estimates of Star-cluster Age, Metallicity, Mass, and Extinction (SESAMME), which we release for public use\footnote{https://github.com/astrolojo/SESAMME}. We describe the structure and computational performance of the code in Section \ref{sec:model}, then apply the code to a suite of synthetic FUV observations of YMCs in Section \ref{sec:validation}. Finally, we further validate the usefulness of the code to real data by analyzing a small sample of four YMCs with published estimates for their ages and stellar metallicities, and show that SESAMME can reliably reproduce past measurements using independent methods. We discuss our validation tests and immediate future uses for the code in Section \ref{sec:discussion}.

\section{The Algorithm}
\label{sec:model}

Given an input spectrum of a star cluster and a set of model spectra for simple stellar populations (SSPs), SESAMME samples the posterior probability distribution function (PDF) for a four-parameter model to return constraints on the physical and observable properties of the cluster. We further describe the necessary inputs and resulting outputs in Sections \ref{subsec:inputs} and \ref{subsec:outputs}.

The first two model parameters -- cluster age and stellar metallicity mass fraction $Z$ -- typically have discrete values determined by the suite of stellar population models used in a SESAMME run. Boundaries on the age and metallicity priors can span the entire range used in a model suite (e.g., ages between 6.0 $<$ log(age) $<$ 11.0 in v2.2 and v2.3 of the Binary Population and Spectral Synthesis (BPASS) models) or can be restricted if a reasonable constraint is already known (e.g., age estimates derived from SED fitting). 

We note that, in the current implementation of SESAMME, the sampler does {\em not} perform any sort of interpolation over the user-specified SSP model grid when generating a model to evaluate. Instead, when the sampler chooses a random value for the cluster age and metallicity, the code will round to the nearest pre-computed value in the model grid. This choice was motivated by the non-linear evolution of some spectral features with respect to age and metallicity in young stellar clusters, which can lead to inaccurate UV line profiles when using a naive interpolation of a coarser model grid. Additionally, the uncertainties on the inferred cluster age and metallicity are typically comparable to or larger than the bin size for each of these parameters (see Tables \ref{tab:synthspec} and \ref{tab:empspec}), making it unnecessary to interpolate to a finer grid.

\begin{table}[tb]
    \caption{{\sc~List of extinction curves available for SESAMME}}
    \vspace{-0.5cm}
    \label{tab:reddening}
    \begin{center}
    \begin{tabular}{ccc}
        \hline
        \hline
       
        Type & Source  & Default $R_V$\\
        \hline
        MW & \citet{fitzpatrick99}  & 3.1\\
         & \citet{ccm89}   & 3.1\\
         & \citet{fm07}  & 3.1\\
         & \citet{Gordon23}  & 3.1\\
        LMC & \citet{gordon03}  & 3.41\\
        SMC & \citet{gordon03}  & 2.74\\
        Starburst & \citet{calzetti2000}  & 4.05\\
        
        \hline

    \end{tabular}
    
    \end{center}
\end{table}

\begin{table*}
\caption{\sc{Example List of Stellar Population Models}}
\vspace{-0.5cm}
    \label{tab:stelmodels}
    \begin{center}
    \begin{tabularx}{0.93\textwidth}{c|P{4cm}P{4cm}P{4cm}}
        \hline
        \hline

        Property & BPASS v2.3 & BPASS v2.2 & Starburst99 \\
        \hline
        Metallicity & [10$^{-5}$, 10$^{-4}$, 0.001, 0.002, 0.003, 0.004, 0.006, 0.008, 0.01, 0.014, 0.02, 0.03, 0.04] & 
        [10$^{-5}$, 10$^{-4}$, 0.001, 0.002, 0.003, 0.004, 0.006, 0.008, 0.01, 0.014, 0.02, 0.03, 0.04] & 
        [0.001, 0.004, 0.008, 0.02, 0.04] \\
        Age range (yr) & 10$^6$ -- 10$^{11}$ & 
        10$^6$ -- 10$^{11}$ & 
        Variable \\
        Age step & 0.1 dex & 0.1 dex & Variable \\
        \hline
        IMF options & Double power law & 
        Single and double power laws; power law with exponential cutoff \citep{chab03} & 
        Double power law\\
        Abundance information & [$\alpha$/Fe] $\in$ [-0.2, 0.0, 0.2, 0.4, 0.6]. \citet{Asplund09} Solar abundances. &
        [$\alpha$/Fe] = 0.0. \citet{Grevesse93} Solar abundances. &  Varies with stellar evolution tracks.\\
        Notes & Includes models with binaries and with single stars only. & Includes models with binaries and with single stars only. & User defines the age range and intervals at which spectra are printed.\\

        \hline

    \end{tabularx}
    
    \tablecomments{Summary of the main characteristics of some frequently-used SSP model suites. This is not a complete list of what SESAMME can accept as the input model suite, but showcases the flexibility of SESAMME to fit the user's preferred set of model assumptions and parameter spaces.}
    \end{center}
\end{table*}

The third parameter is the degree of reddening $E(B-V)$, in magnitudes, which is then translated to extinction $A_V$ assuming a user-specified extinction law. Sampled values of $E(B-V)$ are continuous within a user-specified range. We use the Python packages {\tt dust\_extinction} by K. Gordon\footnote{https://dust-extinction.readthedocs.io/} \citep{GordonDust} and {\tt extinction} by K. Barbary\footnote{https://extinction.readthedocs.io/} \citep{BarbaryExt} to implement the many extinction model options available to SESAMME (see Table \ref{tab:reddening}  for those that are currently implemented). We note that although some of the extinction models in these packages (e.g., the MW-type model from \citet{Gordon23}) allow for variable values of $R_V$, SESAMME assumes that $R_V$ is fixed for a given choice of model. The source code for SESAMME can be easily modified by users who wish to implement additional extinction models or assume alternative values of $R_V$ for existing models.

The final parameter is a continuous rescaling parameter log($A$), which can account for the difference between the assumed stellar mass of the models (e.g., initial mass $10^6 M_{\odot}$ in BPASS) and the actual stellar mass of the cluster under inspection, but {\em only if the flux array of the spectrum is in physical units}. That is, if the input spectrum has been relatively flux calibrated, such that the overall continuum shape and line profiles are correct but the flux array is given in arbitrary units, then log($A$) is a pure unit conversion factor to facilitate comparisons between the data and the model. However, if i) the flux array is fully calibrated and ii) a distance to the target is known, such that the spectrum can be converted to luminosity units before analyzing with SESAMME, then the value of $A$ accounts for the difference between the assumed mass of the models and the actual mass of the cluster.  his means that the mass estimate is thus degenerate with the assumed distance to the cluster, but such a degeneracy is not unique to SESAMME. We note that SESAMME does not explicitly take distance uncertainties into account, which adds an additional systematic source of uncertainty in the log($A$) parameter that is not captured by the modeling process. This means that the errors on log($A$) inferred by SESAMME should be considered lower limits on the true uncertainty. Finally, we note that mass estimates derived using SESAMME are also limited by the completeness of the observations. For example, if a star cluster is more extended than the spectroscopic aperture, the $A$ value inferred by SESAMME will represent a lower limit in mass.

SESAMME has been explicitly tested using the SSP models from BPASS v2.2 \citep[``Tuatara";][]{bpassv22}, BPASS v2.3 \citep[``Broc";][]{bpassv23}, and Starburst99 v7.0 \citep{starburst99} on both synthetic (Sec. \ref{sec:validation}) and empirical (Sec. \ref{sec:cosspec}) ultraviolet spectra of star clusters. The distributed version of SESAMME v1.0 uses BPASS v2.3 with binary stars and a Solar $\alpha$-element abundance [$\alpha$/Fe] = 0 as the default stellar population model set. However, we stress that {\em any} set of stellar population models can be used with SESAMME, as long as the model spectra can be formatted appropriately (see Section \ref{subsec:inputs}). This flexibility allows users to use the SSP suite that is preferred for their science goals and/or cross-examine the impact of different physical assumptions on the age and metallicity constraints inferred by SESAMME. For example, Starburst99 does not take the physics of binary stars into account, while BPASS can model populations that include binary systems, which can significantly affect the time evolution of a stellar population's UV luminosity. Similarly, BPASS v2.3 and v2.2 differ in the assumed stellar atmosphere model grids, the choice of Solar chemical composition, and the available range of options for $\alpha$ element enhancement and initial mass function (IMF). We summarize the major differences between these three SSP model codes in Table \ref{tab:stelmodels}.

Evaluating the posterior probability of a model in SESAMME is straightforward and follows standard practices set out by the authors of \emcee{} \citep{emcee13}. Priors are uniform within boundaries set by the user before the modeling run. The logarithm of the model's likelihood is computed as
\[-\frac{1}{2} \sum_n \frac{(y_{n} - m_{n})^2}{\sigma^2_{n}} + ln(2\pi\sigma^2_{n}),\]
where $y_n$ and $\sigma_n$ are the flux and flux uncertainty arrays of the observed spectrum at wavelength bin $n$, and $m_n$ is the emission model (consisting of the SSP spectrum and an optional nebular continuum component --  see below) generated by the sampler in the same bin. The function that calculates the log-probability also requires a pixel mask (see Section \ref{subsec:inputs}) that removes undesirable wavelength bins from the computation.

\subsection{Nebular Continuum}
\label{subsec:nebular}

When generating samples from the pre-generated SSP model cube (described in Section \ref{subsec:inputs}), SESAMME can also add in a nebular continuum component. Nebular continuum emission can account for a substantial fraction of emission along the line of sight to a star cluster, particularly at young ages and low metallicities. In Section \ref{subsec:synth_nebcont}, we explore how the exclusion of nebular continuum emission in the models affects the inferred properties and their uncertainties.

When nebular continuum emission is ``turned on," the continuum is generated with each sampling, simultaneous with the selection of a stellar population model. Our approach is similar to the subroutine {\tt continuum} in Starburst99. We use the wavelength-dependent emission coefficients given in \citet{ferland80} and \citet{aller84}, which together account for free-free, bound-free, and two-photon emission from H and He$^+$. These values assume typical Case B conditions (densities and temperatures of order $\sim$100 cm$^{-3}$ and $10^4$ K, respectively, as well as an ionizing photon escape fraction of $\sim$0) and a number ratio He/H = 0.1 for the emitting gas. We note that, in this formulation, the intensity and shape of the nebular continuum is agnostic to the gas-phase metallicity.
\looseness=-2

The strength of the nebular continuum emission is determined by the incident ionizing radiation field associated with a given stellar population. For BPASS model spectra, for example, we use the tables of hydrogen-ionizing photon rates $Q$ that are provided in the v2.2 and v2.3 releases. SESAMME then interpolates the computed nebular continua to the wavelength grid of the stellar population models, adds the nebular component to the stellar model, and finally rescales the total model spectrum by the parameter $A$. 
We note that this approach may not be robust for very low mass star clusters, where the high-mass end of the IMF may be sparsely or stochastically sampled \citep{Orozco22}. Lastly, we note that some codes for generating stellar population models already include the nebular continuum in the model spectra \citep[e.g., Starburst99;][]{starburst99}. In these cases, the SESAMME user can turn off the nebular continuum feature.

\subsection{Inputs for SESAMME}
\label{subsec:inputs}

\begin{figure*}[t]
\centering
\includegraphics[width=0.7\linewidth,trim={0.3cm 0.3cm 0.3cm 0.3cm},clip]{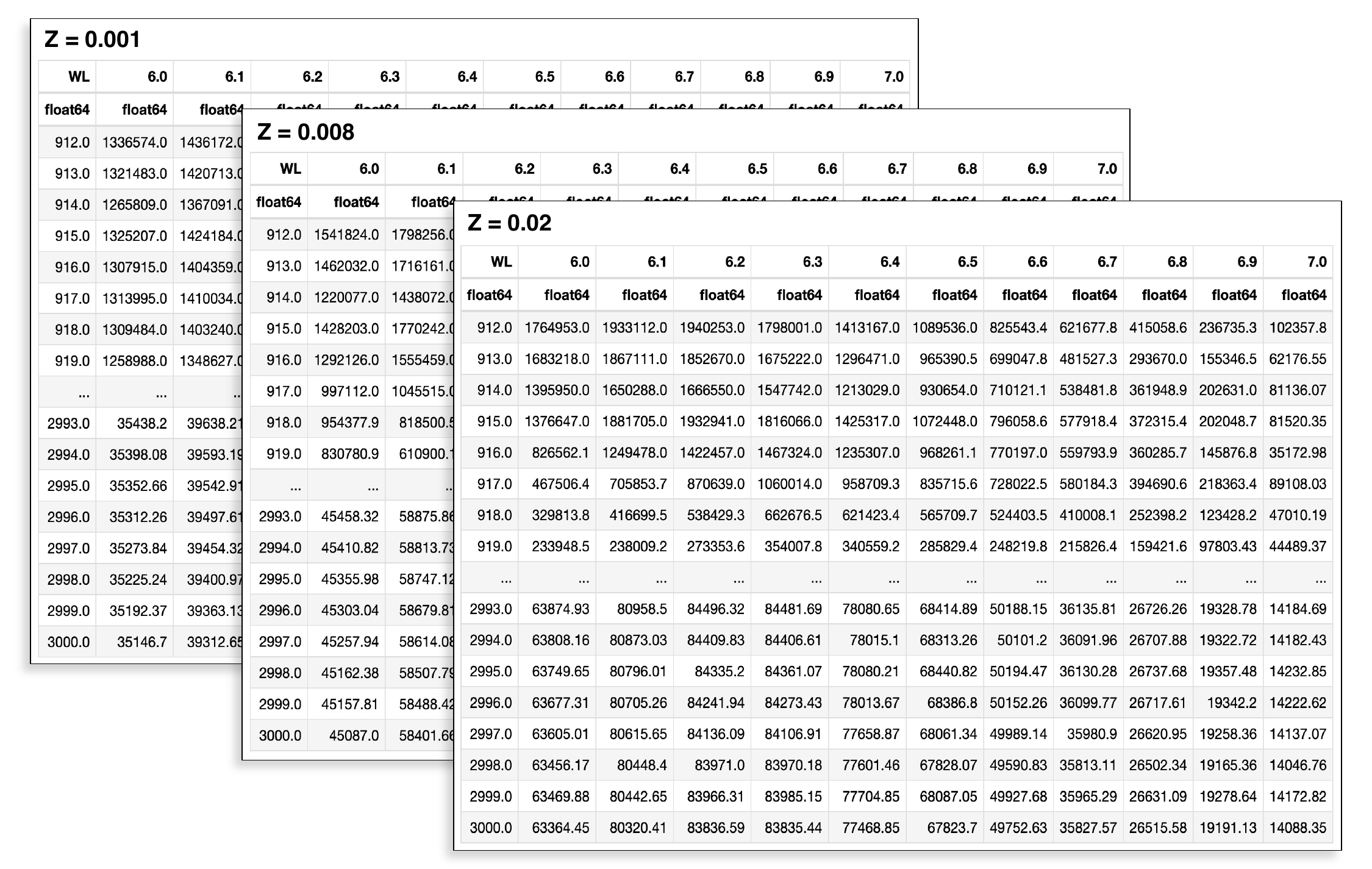}
\caption{
Visualization of the format of a SESAMME model cube, generated from BPASS v2.3 models with binary physics and a solar $\alpha$/Fe, in the form of Astropy Tables. Within each slice of the model cube, the first column gives the wavelength grid (which, in this case, has been truncated to only cover the rest-frame UV), while the remaining columns give the light output at a fixed log(age) at each wavelength bin in units of $L_{\odot}$ \AA$^{-1}$. For clarity, only the first ten age intervals and three of the 13 possible metallicities in BPASS are shown. 
}
\label{fig:modelcube}
\end{figure*}

SESAMME uses \emcee{} \citep{emcee13} for its statistical machinery and requires minimal tuning before it can be run. For example, at the start of every SESAMME run, the user specifies the size of an ensemble of Markov chains or ``walkers," each of which undergo a semi-random walk in the parameter space. In general, we recommend a minimum of $\sim$64 walkers, as we find that a larger ensemble of walkers with moderately long MCMC chains provides better constraints on the PDF than a small ensemble with long chains. 

We also note that, if any of the four parameters have existing measurements from the literature, then the (flat) priors on the relevant parameters can be restricted before running SESAMME. For example, when applying SESAMME to IL spectra of YMCs in M83 in Section \ref{sec:cosspec}, we restrict our age and metallicity priors to a moderately narrow range of values for these parameters, since estimates of the cluster ages and metallicities are known from \citet{Hernandez19}. Finally, the user must also specify which extinction curve to assume throughout the modeling process by setting the parameter {\tt use\_ext\_law}, as well as whether to include a nebular continuum emission component in the modeling process. We list the available extinction curves in Table \ref{tab:reddening} and revisit the topic of nebular continuum emission in Section \ref{subsec:synth_nebcont}.

In addition to setting these pre-run parameters, SESAMME also expects three files as input: 

1) A spectrum in ascii format with three labelled columns for wavelength, flux, and flux uncertainty. For ease of comparison to models, SESAMME assumes that the input spectrum has been corrected for i) redshift/radial velocity and ii) Galactic extinction. The code generally does not require that ISM emission and absorption features be modeled and removed from the spectrum before running SESAMME, as these can be ignored during the modeling process through the use of pixel masks. However, for FUV observations, we recommend that the user models and subtracts the Ly$\alpha$ absorption profile prior to modeling with SESAMME; for very low-redshift clusters, this could include a Milky Way foreground component in addition to Ly$\alpha$ absorption intrinsic to the target. If the Ly$\alpha$ absorption profile is sufficiently broad, the wings can suppress the continuum around and the apparent strength of N{\sc~v}~$\lambda\lambda1238,1242$, which can have small but significant effects on the resulting age and metallicity posterior distributions. 

Additionally, the spectral resolution of the chosen stellar population models may be either too fine or too coarse, depending on the data the user wishes to model (e.g., $R \sim 3000$ optical spectra from Keck/DEIMOS vs. $R \sim 20,000$ FUV spectra from \hstcos{}). We recommend that the user smooths and resamples either the empirical spectrum or the model set, whichever is finer, to the coarser wavelength grid to ensure a proper comparison between models and data. We show an example of this in Section \ref{sec:cosspec}, where we use SESAMME to model {\em HST}/COS FUV spectra (native resolution 0.06 \AA{} per resolution element) with BPASS model spectra (with fluxes given in bins of 1 \AA).

2) A single FITS file (hereafter the ``model cube") containing the set of stellar population models to compare with the empirical spectrum. The model cube is generated by the user with a supplemental piece of Python code (included in the SESAMME release) prior to running the main algorithm for SESAMME. The cube generator collates a set of stellar population models into a single multi-extension FITS file, where each FITS extension contains all model spectra for a given metallicity.  Each extension is formatted as a table in which every column is the model spectrum of a fixed age (except for the first column, which gives the wavelength grid in \AA). By default, model fluxes are assumed to be given in units of $L_{\odot}$ \AA$^{-1}.$ For example, if the user wishes to use BPASS v2.3 models with a given [$\alpha$/Fe] value, the resulting FITS file will have 13 extensions (one per stellar metallicity in BPASS). We illustrate the format of the model cube in Figure \ref{fig:modelcube}, where we render the BPASS v2.3 models with binary physics and a solar [$\alpha$/Fe] value as an Astropy Table. 

The generation of a model cube, rather than using individual model spectra as provided by their respective developers, serves two purposes. First, the cube generator allows the user to truncate the model spectra to only the approximate wavelength regime that is relevant for their analysis (instead of, e.g., the full BPASS spectra which give fluxes from 1 to 100,000 \AA~in steps of 1 \AA). Limiting the wavelength coverage of the model spectra greatly reduces the file size of the model suite and speeds up the MCMC process. Second, collating all model spectra for a given IMF, abundance ratios, or stellar tracks into a single file is faster and simpler to load, manipulate, and examine. This is especially true for users who wish to examine the effects of varying the assumed IMF or abundance ratios when modeling stellar populations, where the number of individual files would otherwise quickly become unwieldy.

3) A {\tt NumPy} boolean array object to act as a pixel mask, which blocks out problematic regions of the spectrum to be modeled (e.g., chip gaps, poorly-subtracted sky lines, or emission and absorption features from Galactic and extragalactic ISM along the line of sight). We recommend creating and saving the pixel mask prior to running SESAMME and loading it as an input file (assumed to be in ascii format).

\subsection{Model Evaluation, Outputs, and Performance}
\label{subsec:outputs}

The basic unit of output from SESAMME is an {\tt EnsembleSampler} object, which is the fundamental vehicle for information in \emcee. Once the MCMC run is complete, the sampler describes the state of all $N$ four-dimensional walkers at each of the $S$ steps in the MCMC chain generated by \emcee{} -- in other words, an array of dimension $4 \times N \times S$. The sampler must then be further processed (using standard methods and functions described in \emcee{}) before further analysis of the posterior PDFs. First, we recommend discarding a number of ``burn-in" steps in which the walkers explore the parameter space to find regions of maximum probability density. The specific number to discard (usually $\sim$a few hundred to 1000) should be determined by the user after examining the completed chain. Further, because consecutive samples in an MCMC chain are not fully independent, many steps may be needed before the ensemble ceases to retain information about past states (that is, to become independent). We recommend ``thinning" the chain by a factor (approximately $0.5 - 1 \times$ the autocorrelation time, which is computed with the \emcee{} function {\tt get\_autocorr\_time}) that must also be determined after completing the MCMC run. Typical autocorrelation times for each variable in SESAMME are of order $\sim$a few hundred steps. We find that using a thinning factor that is roughly equal to the average of the four autocorrelation times is usually sufficient to achieve a good sampling.

From the validation tests using either synthetic or empirical UV spectra that we describe in Sections \ref{sec:validation} and \ref{sec:cosspec}, we find that the sampler usually settles onto its preferred solution(s) relatively quickly, with only small gains in precision when using very long chains ( $> 50,000$). Chains as short as 10,000 steps are often sufficient for the walker ensemble to stabilize around regions of maximum density, even if formal convergence is not reached (as determined by estimates of the autocorrelation time; see the next section).

SESAMME's computational performance is adequate for use on personal computers. For reference, most of the development and testing of SESAMME v1.0 was done on a laptop with a 2.3 GHz Intel Core i9 processor and 32 GB of DDR4 memory. Using this setup, \emcee{} runs at $\sim$5 iterations per second with the nebular continuum generation turned on (or $\sim$15 iterations per second with no nebular continuum); a single run of 50,000 steps takes approximately 150 minutes (50 minutes with no nebular continuum) to complete without CPU parallelization.

\section{Validating the Algorithm}
\label{sec:validation}

\subsection{Testing on synthetic spectra}
\label{subsec:synthtest}

To illustrate the extent to which SESAMME can recover accurate stellar ages and metallicities even from spectra with moderate signal-to-noise ratio S/N, we created a suite of mock cluster IL spectra based on models from BPASS v2.3. We used the BPASS v2.3 binary models with [$\alpha$/Fe] = 0, beginning with the pure stellar population model of a given age, $Z$, and mass. Then we added an appropriately scaled nebular continuum component and reddened the combined stellar+nebular spectrum assuming a \citet{ccm89} extinction law with $R_V = 3.1$. Because our initial science applications with SESAMME will involve far-UV spectra of YMCs from {\em HST}/COS, the eight mock spectra that we generate span 1 dex in age, metallicity, and reddening, but are generally fairly young (at most 30 Myr). The true model parameters for the eight mock spectra are given in columns 2 -- 5 of Table \ref{tab:synthspec}. All spectra assume a broken power law IMF with low-mass slope $\alpha_1$ = -1.30, high-mass slope $\alpha_2$ = -2.35, and a maximum mass of 300 $M_{\odot}$ (``135\_300" in BPASS v2.2 and v2.3).

We then added a small noise component to the flux in each wavelength bin to ``jitter" the spectrum away from an ideal stellar population model. The noise amplitude per wavelength bin (which is distinct from the flux {\em uncertainty} in that bin) was drawn from a uniform distribution such that the flux is changed by up to $\pm$15\% from its noiseless value. We similarly created a flux uncertainty array by drawing from a normal distribution, scaled to the noiseless model flux in each bin, such that the typical flux uncertainty is 30 $\pm$ 10\%. We then convolve this flat noise profile with the throughput of the combined COS FUV G130M+G160M gratings, which creates a noise array with similar shape to a typical observation with COS. The final noise array is scaled such that the S/N ranges from $\sim$2 to 15 per \AA{} across the 1150 -- 1800 \AA{} range we use to approximate the COS wavelength coverage. Finally, we create a pixel mask that blocks out the approximate locations of COS chip gaps, as well as narrow wavelength ranges ($\Delta$\AA{} $\sim$5 -- 20) at the expected locations of strong ISM emission and absorption features (e.g., OI~$\lambda$1302, CII~$\lambda$1334, SiIV~$\lambda\lambda$1393,1402, CIV~$\lambda\lambda$1548,1550, and geocoronal Ly$\alpha$). 

Columns 6 -- 9 of Table \ref{tab:synthspec} show the 16, 50, and 84$^{th}$ percentiles of the samples in each of the four marginalized distributions. SESAMME is generally successful in recovering accurate values for all 4 parameters across the parameter space that we explore in this work, with $\sim$10 -- 15\% uncertainties on the cluster age and mass, stellar metallicity, and average reddening.

\begin{table*}[tb]
    \caption{{\sc~Input and recovered properties of synthetic validation spectra$^a$}}
    \label{tab:synthspec}
    \begin{tabular}{c|cccc|cccc}
        \hline
        \hline
        & \multicolumn{4}{c|}{Input parameters} & \multicolumn{4}{c}{PDF medians from SESAMME}\\
        \hline
        Name & log(Age) & $Z$ & $E(B-V)^b$ & Mass & log(Age) & $Z$ & $E(B-V)$ & Mass\\
        & (yr) &  & (mag) & (10$^5 M_{\odot}$)& (yr) &  & (mag) & (10$^5 M_{\odot}$)\\
        \hline
        
        1 & 6.5 & 0.002 & 0.1 & 1.00 & 
        6.49$^{+0.10}_{-0.10}$ & 0.002$^{+0.001}_{-0.001}$ & 0.10$^{+0.01}_{-0.01}$ & 0.94$^{+0.34}_{-0.23}$\\
        2 & 6.5 & 0.002 & 1.0 & 1.00 & 
        6.46$^{+0.08}_{-0.08}$ & 0.002$^{+0.001}_{-0.001}$ & 1.00$^{+0.01}_{-0.01}$ & 0.85$^{+0.19}_{-0.19}$\\
        3 & 7.5 & 0.002 & 0.1 & 1.00 & 
        7.48$^{+0.07}_{-0.09}$ & 0.002$^{+0.001}_{-0.001}$ & 0.09$^{+0.02}_{-0.02}$ & 0.97$^{+0.15}_{-0.13}$\\
        4 & 7.5 & 0.002 & 1.0 & 1.00 & 
        7.50$^{+0.10}_{-0.11}$ & 0.002$^{+0.001}_{-0.001}$ & 1.01$^{+0.02}_{-0.02}$ & 1.08$^{+0.19}_{-0.17}$\\

        5 & 6.5 & 0.020 & 0.1 & 1.00 & 
        6.51$^{+0.04}_{-0.04}$ & 0.020$^{+0.004}_{-0.003}$ & 0.10$^{+0.02}_{-0.01}$ & 1.00$^{+0.15}_{-0.14}$\\
        6 & 6.5 & 0.020 & 1.0 & 1.00 & 
        6.54$^{+0.08}_{-0.06}$ & 0.020$^{+0.003}_{-0.002}$ & 1.01$^{+0.01}_{-0.01}$ & 1.16$^{+0.31}_{-0.19}$\\
        7 & 7.5 & 0.020 & 0.1 & 1.00 & 
        7.47$^{+0.06}_{-0.08}$ & 0.020$^{+0.004}_{-0.003}$ & 0.11$^{+0.03}_{-0.02}$ & 1.01$^{+0.11}_{-0.11}$\\
        8 & 7.5 & 0.020 & 1.0 & 1.00 & 
        7.49$^{+0.05}_{-0.08}$ & 0.022$^{+0.004}_{-0.003}$ & 1.00$^{+0.02}_{-0.01}$ & 0.92$^{+0.11}_{-0.11}$\\

        \hline

    \end{tabular}
    \begin{center}
    \tablecomments{$^a$ All use the BPASS v2.3 model with a broken power law IMF\\ (``135\_300" in BPASS v2.2 and v2.3) and solar levels of $\alpha$/Fe.\\
    $^b$ Values are equivalent to $A_V$ = 0.31 and 3.1 for the \citet{ccm89} extinction law with $R_V = 3.1$.}
    \end{center}
\end{table*}

One of the key caveats of the L12 technique is that its accuracy in determining stellar metallicities requires an age-accurate choice of isochrone, especially for very young clusters whose properties evolve significantly over the course of $\sim$a few Myr. Indeed, avoiding such age assumptions was one of the driving forces behind the creation of SESAMME. However, the code and user must make several assumptions which may affect the posterior PDF for a given SESAMME run. In the following sections, we explore the sensitivity of SESAMME's output to changes in the assumed extinction curve and in the assumed prevalence of nebular continuum emission. 

\subsection{Varying the Extinction Law}
\label{subsec:synth_ext}

The eight model spectra we describe above were reddened assuming a Milky Way-like extinction curve with $R_V$ = 3.1, and for simplicity our nominal runs with SESAMME assume the same curve in the modeling process (i.e., with the parameter {\tt use\_ext\_law} set to the \citet{ccm89} curve). To examine the impact of an improper choice of extinction curve on the inferred stellar properties, we also ran the eight synthetic spectra through SESAMME with {\tt use\_ext\_law} set to the \citet{calzetti2000} starburst extinction curve with $R_V$ = 4.05. All other parameters within SESAMME (e.g., priors, chain lengths, etc.) were identical to the baseline runs. 

When using the \citet{calzetti2000} law to model spectra that, in reality, are subject to Milky Way-like extinction, SESAMME infers reddening values that are higher than the true values by as much as $\sim30$\%, and the inferred cluster mass is also overestimated by 0.2-0.3 dex on average. The median ages increase by 0.1 dex at most, and the median metallicities are essentially unaffected. Further, upon visually comparing the models that correspond to the median PDF values in both the MW and the Calzetti case, it becomes clear that assuming a Calzetti-like extinction curve provides an inferior ``best fit" to our synthetic spectra. For example, for spectrum 5 (young, high-metallicity, and $E(B-V)$ = 1), the optimal model with MW-like reddening yields a reduced $\chi^2 = 0.97$, while the optimal model with the Calzetti curve yields $\chi^2 = 4.9$. These changes are driven by a combination of i) the shallower slope at very blue wavelengths in the \citet{calzetti2000} curve compared to the \citet{ccm89} curve,  and ii) the lower S/N at the bluest and reddest wavelengths of our synthetic spectra (such that the model fitting is heavily weighted towards the highest S/N regions of the spectrum around $\sim 1500$ \AA). These factors lead to a situation where the optimal model is more heavily reddened (to better approximate the overall continuum shape), but correspondingly more massive to compensate for the dearth of flux at all wavelengths that results from the increased reddening. This test highlights the importance of accurately selecting an extinction curve that best describes the specific environmental conditions of the targets.

\subsection{Effect of the Nebular Continuum}
\label{subsec:synth_nebcont}

We remind users of SESAMME v1.0 that SESAMME's default behavior is to add nebular continuum emission to the stellar component during the modeling process. However, for flexibility, we also allow users to turn nebular continuum emission off.

\begin{figure}[t]
\centering
\includegraphics[width=\linewidth,trim={0.25cm 0.25cm 0.2cm 0.2cm},clip]{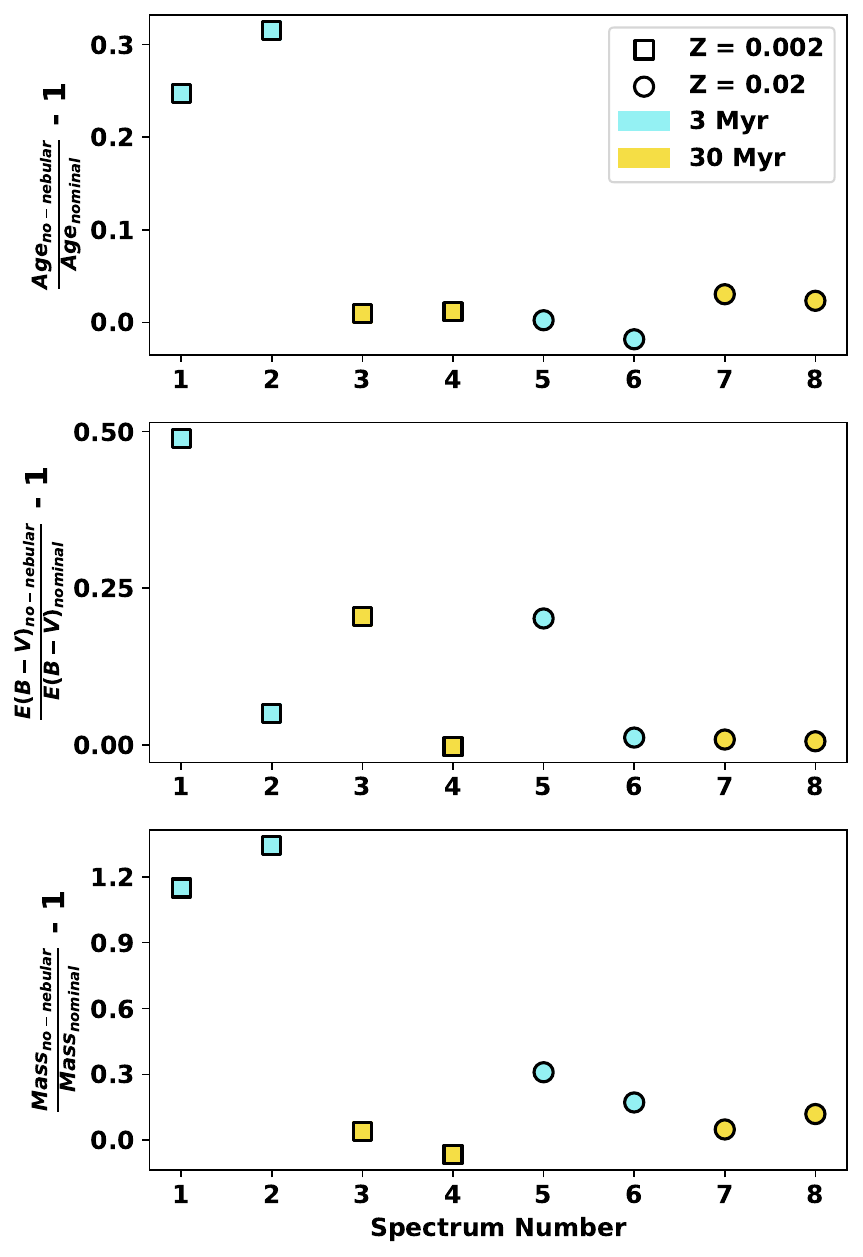}
\caption{
Fractional residual difference of the ages (top), reddening (middle), and masses (bottom) inferred for the eight synthetic FUV spectra with (``nominal") and without (``no-nebular") a nebular continuum component in the modeling. All points are colored according to the input age for that mock cluster spectrum (blue for 3 Myr, gold for 30 Myr). Squares denote low-metallicity clusters (spectra 1 -- 4) and circles denote solar-metallicity clusters (spectra 5 -- 8). 
}
\label{fig:noneb}
\end{figure}

To illustrate how the (non)inclusion of nebular continuum emission affects the inferred properties of a star cluster, we re-ran the eight synthetic spectra through SESAMME with the nebular continuum component set to zero at all wavelengths during the modeling process (hereafter the ``no-neb" case), with all other parameters identical to the baseline run. In Figure \ref{fig:noneb}, we compare the median ages, reddening values, and masses that are inferred from the baseline and no-neb runs; the median metallicity does not change significantly between the baseline and no-neb cases and thus is not shown in the figure. From these tests, we can conclude that the non-inclusion of nebular continuum emission in the models has the following effects on each of the four variables:

i) Age -- Cluster ages in the no-neb case are slightly but systematically older than in the baseline case, with the most extreme differences for very young and metal-poor clusters (clusters 5 and 6).

ii) Metallicities -- Differences in the inferred metallicities are negligible within the uncertainties. However, the metallicity uncertainties seem to broaden slightly at low $Z$ values when nebular continuum is not taken into account.

iii) Reddening -- SESAMME prefers models that are systematically more reddened in the no-neb case than in the baseline case, with differences as large as 50\% for the youngest and most metal-poor clusters.

iv) Mass -- For clusters younger than $\sim$10 Myr, SESAMME infers stellar masses that are systematically higher (by up to a factor of 2) in the no-neb case.

These behaviors are mostly expected due to the nature of nebular continuum emission and how SESAMME treats its inclusion in the modeling. SESAMME's default modeling procedure includes a physically motivated amount of nebular continuum emission -- that is, for a given age, mass, and metallicity, the stellar ionizing photon output $Q$ and hence nebular continuum strength is fixed. In the FUV, the nebular continuum rises rapidly from 912 \AA{}, peaks around 1500 \AA, then gently decreases to redder wavelengths. For young stellar populations, this means that the nebular contintuum contributes a greater fraction of UV light (relative to the stellar continuum) at longer wavelengths. The result is a spectrum whose UV slope is shallower than one with purely stellar emission, which can mimic the combined effects of extinction and aging. In situations where nebular continuum is actually present in the system under observation but is omitted from the modeling procedure, the sampler will converge to a model that is both more massive (to compensate for the dearth of flux at the red end) and more reddened (to account for the flatter-than-expected spectral slope). In short, we can conclude that the inclusion of a nebular continuum emission component in the models has a non-negligible impact on the inferred properties of young, massive star clusters and should not be ignored. 

\begin{figure}[t]
\centering
\includegraphics[width=\linewidth,trim={1cm, 0.1cm, 1cm, 0.1cm}, clip]{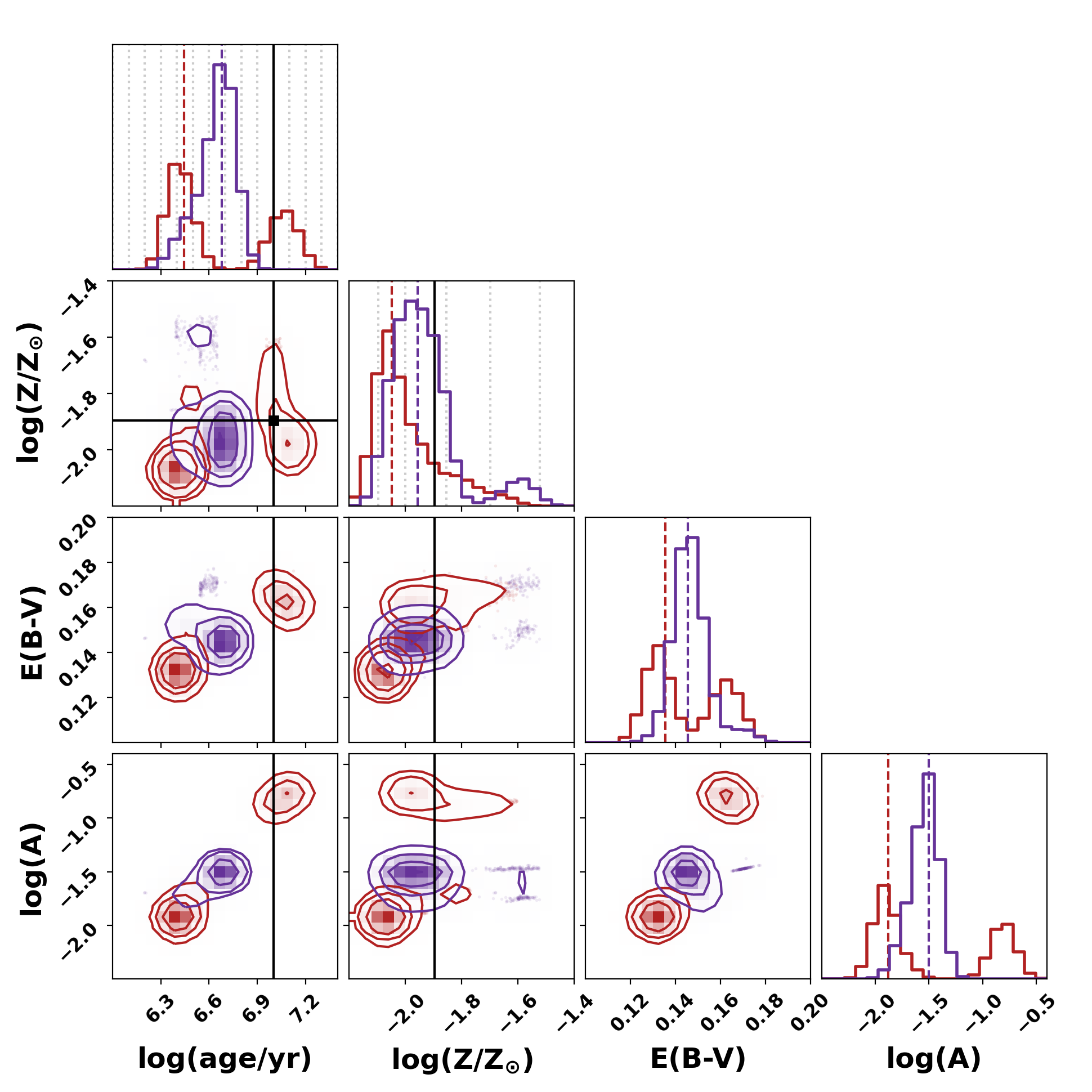}
\caption{
Corner plot for the SESAMME fit to the UV spectrum of M83-8, using either Starburst99 (purple) or BPASS v2.3 (red) as the assumed model suite. This spectrum has a S/N at 1310 \AA{} of $\sim$6.8 at the native 0.06 \AA{}/resel resolution of COS (S/N $\sim$26.8 after binning to at the 1 \AA{} resolution of BPASS). Red and purple dashed lines mark the median of each parameter's PDF for the two model sets. Grey dotted lines mark the BPASS age and metallicity grid points; grid points for our SB99 models are omitted for clarity. Black solid lines mark the best-fit age and metallicity from \citet{Hernandez19}. The 2D and 1D histograms have been lightly smoothed for clarity.
\looseness = -2
}
\label{fig:cornerplot}
\end{figure}

\section{Re-examining YMCs in M83 and NGC 1313}
\label{sec:cosspec}

\begin{figure*}[t]
\centering
\includegraphics[width=\linewidth,height=0.55\linewidth,trim={0.1cm 0.2cm 0.1cm 0.2cm},clip]{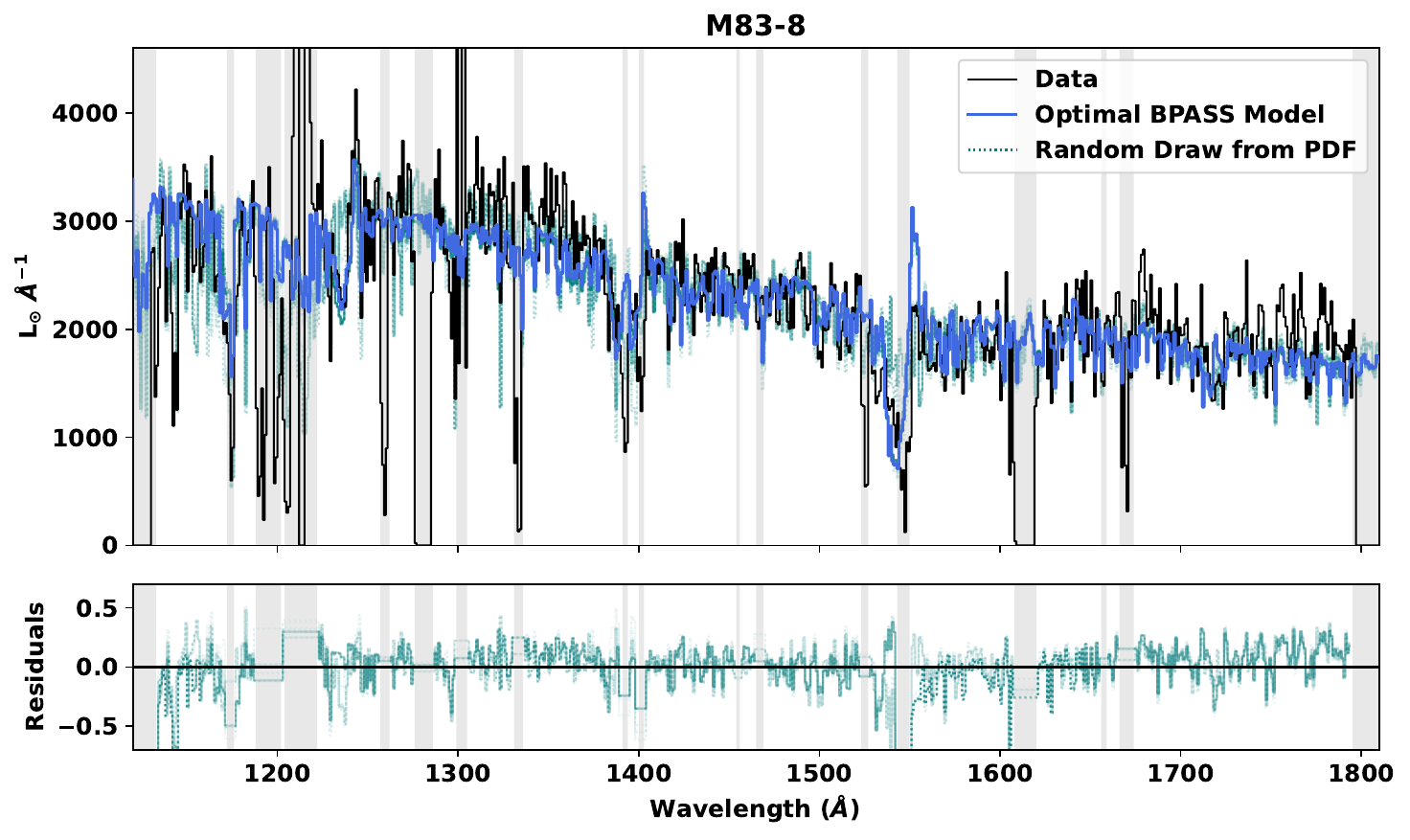}\\
\includegraphics[width=0.4\linewidth,height=0.4\linewidth,trim={0.2cm 0.1cm 0.22cm 0.0cm},clip]{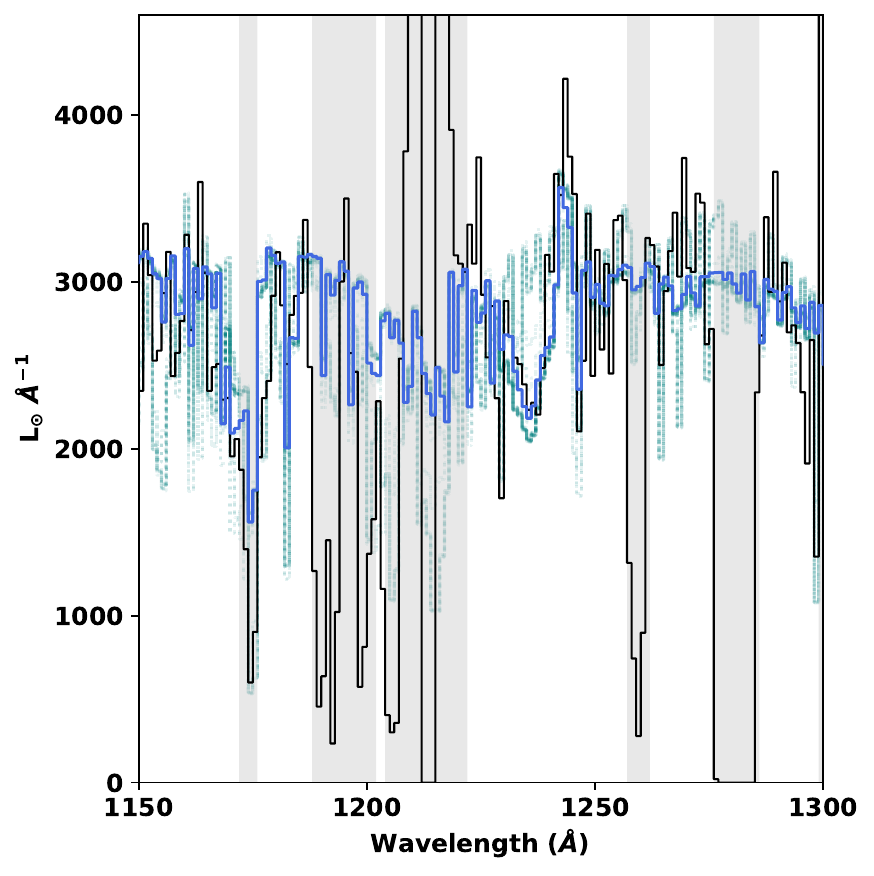}
\hspace{0.5cm}
\includegraphics[width=0.4\linewidth,height=0.4\linewidth,trim={0.2cm 0.1cm 0.22cm 0.0cm},clip]{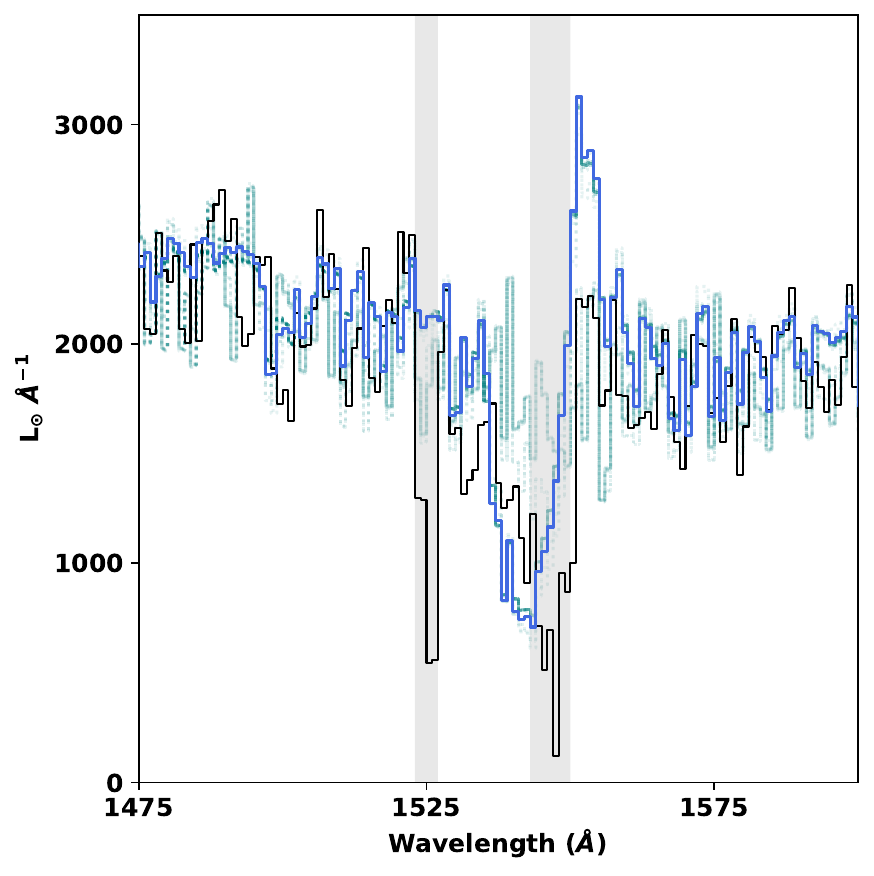}
\caption{
An example of SESAMME's performance in modeling the FUV spectrum of a metal-rich YMC. 
{\em Top:} The black curve shows the H{\sc i} absorption-corrected, combined G130M+G160M spectrum of YMC M83-8 \citep{Hernandez19} from {\em HST}/COS, after smoothing and resampling to the 1 \AA{} resolution of the BPASS models. Intervals in grey shaded regions were masked during the modeling process and include ISM emission and absorption features, chip gaps, and geocoronal Ly$\alpha$ and O{\sc i} {\bf $\lambda 1302$}. The blue curve shows the median model (see text for details), while the teal dotted curves show 50 random samples from the cleaned MCMC chain. {\em Middle:} Model residuals for the 50 samples. {\em Bottom:} Zoom-ins to regions that include age-sensitive features: N{\sc v} {\bf $\lambda 1240$} in the bottom left and C{\sc iv} {\bf $\lambda 1550$} in the bottom right.
}
\label{fig:M83fit}
\end{figure*}

\begin{figure*}[t]
\centering
\includegraphics[width=\linewidth,height=0.55\linewidth,trim={0.1cm 0.2cm 0.1cm 0.2cm},clip]{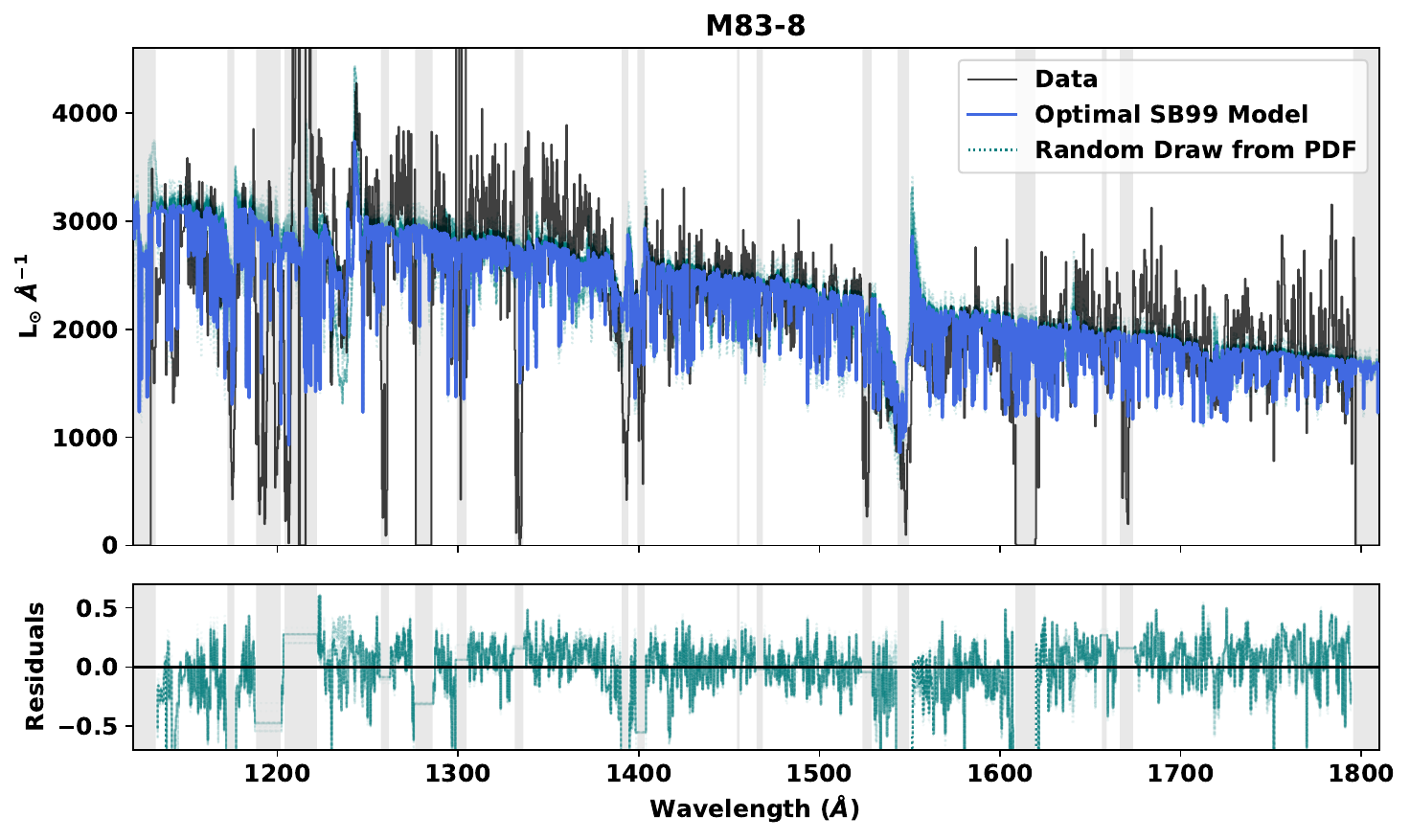}\\
\includegraphics[width=0.4\linewidth,height=0.4\linewidth,trim={0.2cm 0.1cm 0.22cm 0.0cm},clip]{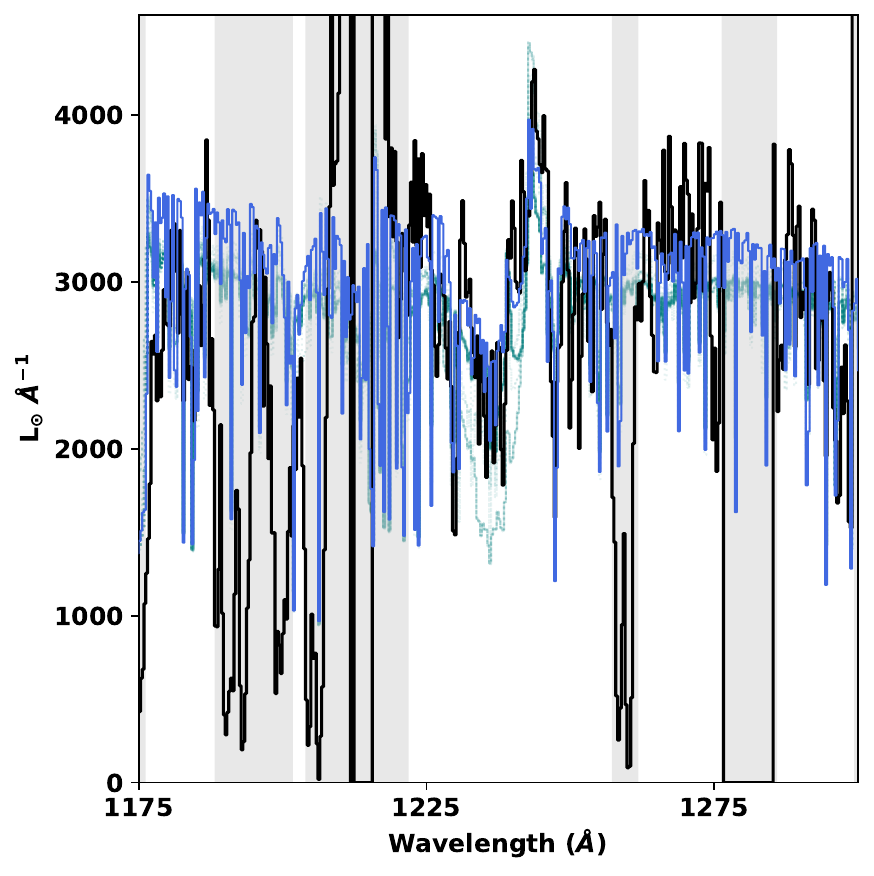}
\hspace{0.5cm}
\includegraphics[width=0.4\linewidth,height=0.4\linewidth,trim={0.2cm 0.1cm 0.22cm 0.0cm},clip]{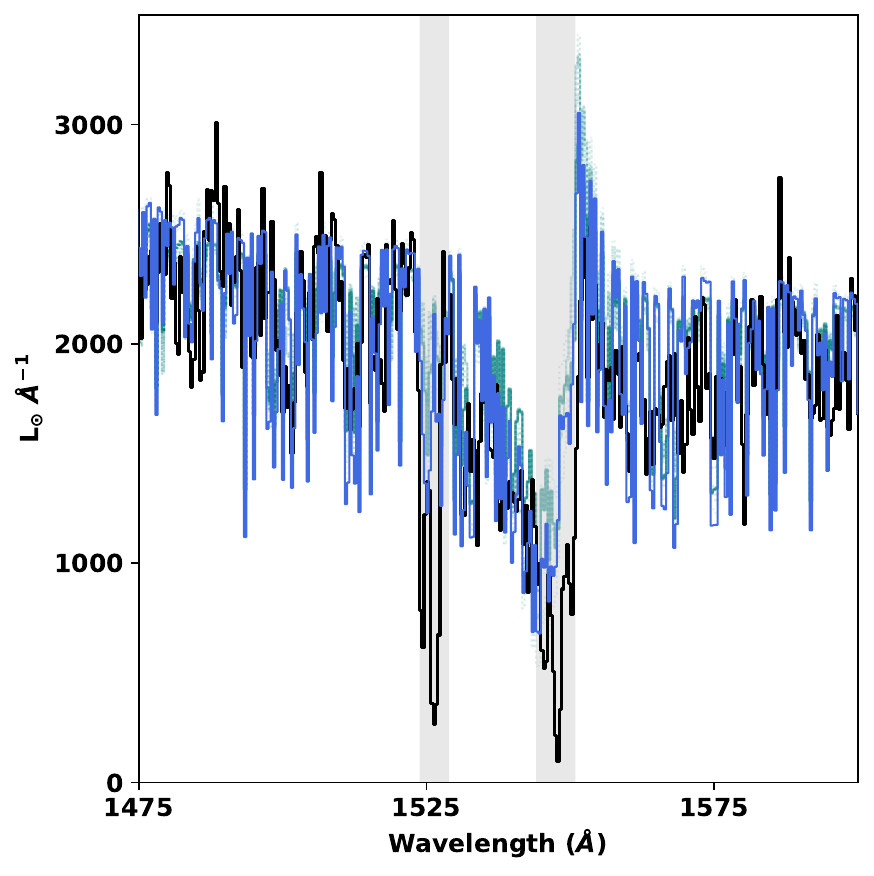}
\caption{
An example of SESAMME's performance using SB99 models to characterize the FUV spectrum of a YMC.
{\em Top:} The black curve shows the H{\sc i} absorption-corrected, combined G130M+G160M spectrum of YMC M83-8 \citep{Hernandez19} from {\em HST}/COS, now smoothed and resampled to the 0.6 \AA{} resolution of our SB99 models. As in Figure \ref{fig:M83fit}, intervals in grey shaded regions were masked during the modeling process, and teal dotted curves show either 50 random samples from the MCMC chain ({\em top}) or the residuals of those samples from the observed spectrum ({\em middle}). In the bottom panels we show zoom-ins to age-sensitive features (N{\sc v} {\bf $\lambda 1240$} on the left and C{\sc iv} {\bf $\lambda 1550$} on the right).
}
\label{fig:M83fit_sb99}
\end{figure*}

As a first science use of SESAMME, we also applied it to a preliminary sample of four FUV spectra of YMCs from {\em HST}/COS. The sample includes two sightlines towards YMCs in the massive, metal-rich spiral M83 \citep[e.g.,][]{Bresolin16} and two towards YMCs in the LMC-metallicity galaxy NGC 1313 \citep[e.g.,][]{walsh97}. For all clusters, we use the combined G130M+G160M data from COS. The spectra towards M83 were originally obtained as part of {\em HST} program 14681 (PI: Aloisi), while the spectra towards NGC 1313 were obtained as part of {\em HST} program 15627 (PI: Adamo).

For each YMC, we created both a high- and low-resolution version of each COS spectrum. The high-resolution spectrum was used only for determining the recession velocity of the cluster, while the low-resolution one was used as input to SESAMME. We downloaded individual spectra for each object from the Mikulski Archive for Space Telescopes (MAST) and reduced them using the calibration pipeline {\sc~CALCOS} v3.4.0. We further processed the individual exposures using an IDL code by the COS Guaranteed Time Observer Team \citep{DanforthCOS}, which interpolates each spectrum onto a common wavelength grid before weight-combining them. Lastly, we rebinned the final spectra by one COS resolution element (1 resel = 6 pixels) that corresponds to the nominal point-spread function. This process also rescales and stitches together the spectra from the different COS gratings into a single, coadded file. These high-resolution spectra have a typical S/N of $\sim$5 -- 7 per resel (0.06 \AA~for G130M and 0.07 \AA~for G160M) at 1310 \AA.
\looseness=-2

We correct for Galactic reddening along the line of sight using the extinction maps by \citet{Schlafly11}, assuming a \citet{ccm89} extinction curve with $R_V = 3.1$. We also fit and divide out a Ly$\alpha$ absorption profile, which includes a foreground component from Milky Way H{\sc~i}, using the Python software {\tt VoigtFit} \citep{Voigtfit18}. We convolve the COS line spread function profiles with the FWHM of the source in the dispersion direction (as measured from the acquisition images), which accounts for instrument-induced broadening of the Ly$\alpha$ absorption profile. This procedure is similar to that used in \citet{James14} and \citet{Hernandez20,Hernandez21}.

We then use the high-resolution spectrum to measure the recession velocity of the cluster from photospheric lines. After fitting the local stellar continuum around the relevant lines, we fit a series of Gaussians to 4 -- 6 photospheric lines across the wavelength range of the spectrum; these typically included  C{\sc~iii}~$\lambda1247$, Si{\sc~iii}~$\lambda1294$, C{\sc~ii}~$\lambda1324$,  O{\sc~iv}~$\lambda1342$, and S{\sc~v}~$\lambda1502$. We take the average redshift of these lines as the recession velocity of the cluster and use it to shift the COS spectrum to the rest frame.

Proper comparisons with stellar population models must account for the differing spectral resolutions of the COS spectra and the models (e.g., 1 \AA{} for BPASS). To do this, we smooth the spectrum with a Gaussian filter, such that $\sigma_s = \sqrt(\sigma_{model}^2 - \sigma_{COS}^2)$. We then resample the smoothed spectrum onto the model wavelength grid (e.g., in steps of 1 \AA{} to match the BPASS wavelength grid). Finally, we use the dereddened, H{\sc~i}-corrected, and smoothed spectra as input to SESAMME.

We placed loose constraints on the age priors according to literature age estimates for each cluster. For the YMCs in M83, the ages and age uncertainties are derived from SED fitting \citep{Hernandez19} and have a typical uncertainty of 50\%. For example, \citet{Hernandez19} estimated the cluster M83-3 to be $5 \pm 3$ Myr old from {\em HST}/WFC3 photometry; therefore, when modeling this cluster with SESAMME, we constrain the age to be at most 10 Myr old. For the YMCs in NGC 1313, the ages and age uncertainties are taken from the CLusters in the UV as EngineS survey \citep[CLUES;][]{Sirressi22}, which uses a two-SSP spectral fit to constrain the properties of the clusters. We use their light-weighted ages as the comparison point for our work. Priors on the stellar metallicity were determined in a similar manner and generally spanned a range of 0.5 -- 2$\times$ the literature value. The reddening and amplitude parameters are left essentially unconstrained.

\begin{table*}[tb]
    \caption{{\sc~Inferred properties of YMCs from HST/COS FUV spectra$^a$}}
    \label{tab:empspec}
    \begin{tabular}{c|cccc|cccc}
        \hline
        \hline
        & \multicolumn{4}{c|}{BPASS} & \multicolumn{4}{c}{Starburst99} \\
        \hline
        Name & log(Age) & $Z$ & $E(B-V)^a$ & Mass & log(Age) & $Z$ & $E(B-V)^a$ & Mass\\
        & (yr) &  & (mag) & (10$^4 M_{\odot}$)
        & (yr) &  & (mag) & (10$^4 M_{\odot}$)\\
        \hline
        M83-3 & 6.53$^{+0.43}_{-0.06}$ & 0.023$^{+0.007}_{-0.002}$ & 0.27$^{+0.02}_{-0.00}$ & 71$^{+417}_{-3}$
        & 6.48$^{+0.09}_{-0.06}$ & 0.023$^{+0.005}_{-0.003}$ & 0.25$^{+0.02}_{-0.00}$ & 51$^{+47}_{-1}$\\
        M83-8 & 6.46$^{+0.59}_{-0.08}$ & 0.010$^{+0.005}_{-0.002}$ & 0.14$^{+0.03}_{-0.01}$ & 2$^{+14}_{-1}$
        & 6.68$^{+0.04}_{-0.14}$  & 0.011$^{+0.003}_{-0.001}$ & 0.15$^{+0.06}_{-0.003}$ & 3$^{+0}_{-1}$\\
        NGC 1313-1 & 6.45$^{+0.606}_{-0.07}$ & 0.005$^{+0.002}_{-0.001}$ & 0.18$^{+0.02}_{-0.01}$ & 5$^{+56}_{-1}$
        & 6.50$^{+0.28}_{-0.08}$ & 0.003$^{+0.002}_{-0.001}$ & 0.12$^{+0.01}_{-0.00}$ & 3$^{+3}_{-0}$\\
        NGC 1313-2 & 6.80$^{+0.38}_{-0.11}$ & 0.003$^{+0.001}_{-0.000}$ & 0.28$^{+0.05}_{-0.05}$ & 8$^{+24}_{-3}$
        & 6.73$^{+0.28}_{-0.05}$  & 0.004$^{+0.000}_{-0.000}$  & 0.28$^{+0.01}_{-0.01}$ & 5$^{+9}_{-0}$\\
        \hline

    \end{tabular}
    \begin{center}
    \tablecomments{$^a$Assumes a \citet{calzetti2000} extinction curve with $R_V$ = 4.05.}
    \end{center}
\end{table*}

We then allow SESAMME to explore the BPASS parameter space using an ensemble of 128 walkers in chains of length 5000 steps, with the nebular continuum component turned on. For each cluster, we use the BPASS v2.3 binary models with solar [$\alpha$/Fe] and a broken power law IMF with low-mass slope $\alpha_1$ = -1.30, high-mass slope $\alpha_2$ = -2.35, and a maximum mass of 300 $M_{\odot}$. We further assume a \citet{calzetti2000} extinction curve, since test runs using other extinction curves generally produce a poorer fit to the data, particularly in the heavily star-forming environments we are interested in. To generate the posterior PDFs for our four parameters, we use \emcee{}'s helper function {\tt get\_chain} to discard the first 300 -- 800 steps (the ``burn-in") of the chain, then thin the chain by a factor of $\sim$100 -- 300. The exact value of the thinning factor is determined at the end of each run using the \emcee{} function {\tt get\_autocorr\_time}. This leaves a collection of several thousand independent samples from which we can construct posterior PDFs for each of the four variables in the model. For each parameter, we take the PDF median as the ``best-fit" value and the 16th and 84th percentile values as the 1$\sigma$ uncertainties. We show the 2D and 1D posterior PDFs as a corner plot in Figure \ref{fig:cornerplot}.

We show an example of SESAMME's performance on the YMC M83-8 in Figure \ref{fig:M83fit}. In addition to the smoothed and rebinned COS FUV spectrum, we also show the ``median model" inferred by SESAMME. This is the total model (stellar plus nebular continuum) for which the four parameters are set to the 50th percentiles of their respective, one-dimensional posterior PDFs (see columns 2 -- 5 of the second row of Table \ref{tab:empspec}). Further, we show 50 random draws from the ensemble sampler as teal curves, which illustrate the range of models that SESAMME considers to be a plausible fit to the data within the uncertainties. We show the median models for the other three clusters in the Appendix (Fig. \ref{fig:appendix}).

To test the self-consistency of SESAMME across different model sets, we then performed this entire procedure a second time using a different set of stellar population models. Specifically, we used Starburst99 single-burst models of a fixed mass of 10$^6 M_{\odot}$, assuming the high mass-loss Geneva stellar tracks and a double-power law IMF with low-mass slope $\alpha_1$ = -1.30, high-mass slope $\alpha_2$ = -2.3, and a maximum mass of 100 $M_{\odot}$. We generated models with $Z$ = 0.001, 0.004, 0.008, 0.02, and 0.04, each of which includes model spectra in steps of 1 Myr for ages between 1 and 100 Myr. Because we are interested in modeling FUV spectra from COS, we used only the high-resolution, theoretical UV spectra ({\tt ifaspec} files; $\sim$0.6 \AA/pix) from Starburst99. After smoothing and rebinning the COS spectra to the 0.6 \AA/pixel wavelength grid of the models, we re-ran the analysis described above for each YMC. The {\tt ifaspec} spectra already include a nebular continuum component in addition to the stellar-only light, so we turned off the nebular continuum feature within SESAMME. All other aspects of each MCMC run (priors, chain length, etc.) were identical to the runs with BPASS models. We show the median SB99 model and 50 random draws from the ensemble sampler in Figure \ref{fig:M83fit_sb99}.

\begin{figure*}[t]
\centering
\includegraphics[height=0.45\linewidth,width=0.45\linewidth,trim={0.25cm 0.2cm 0.2cm 0.2cm},clip]{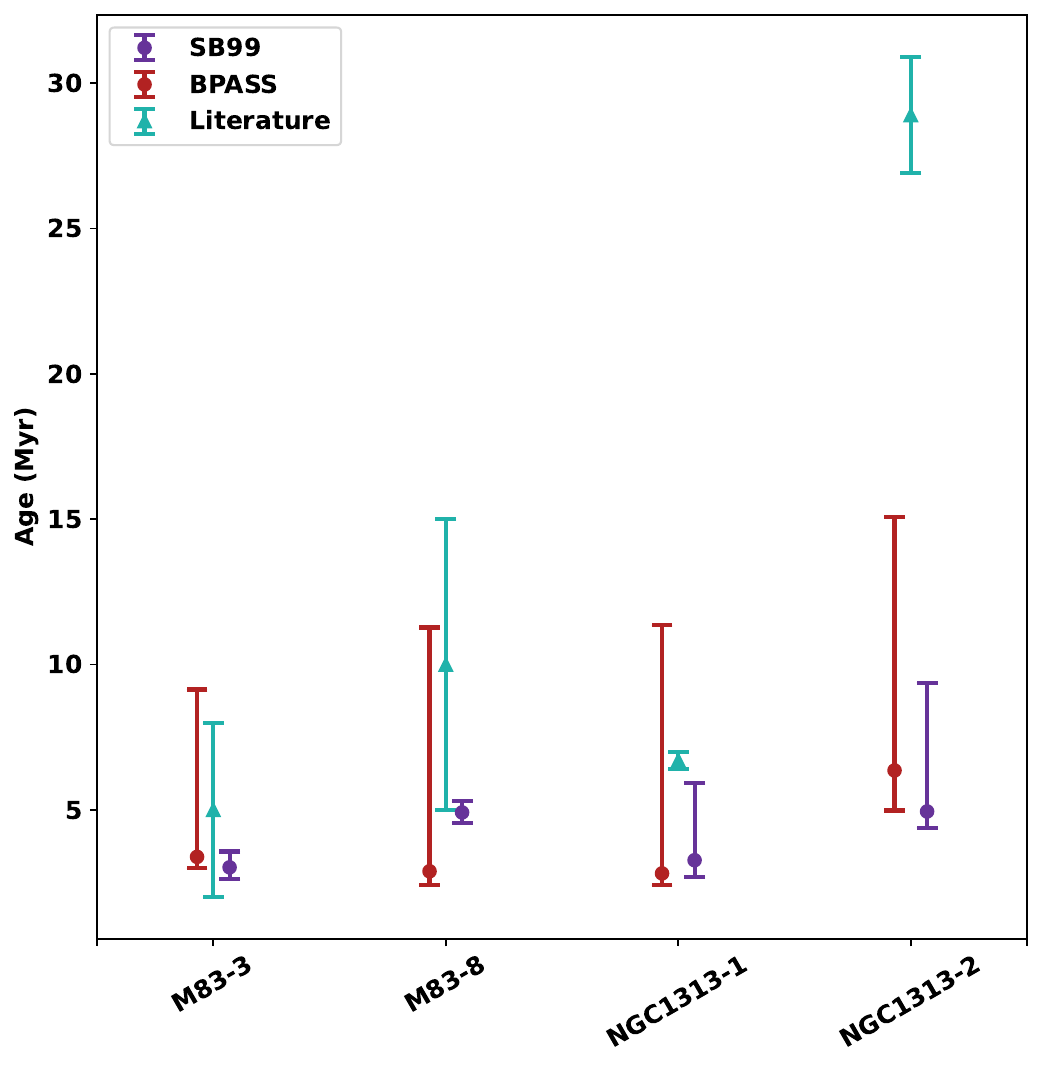}
\hspace{0.5cm}
\includegraphics[height=0.45\linewidth,width=0.45\linewidth,trim={0.25cm 0.2cm 0.2cm 0.2cm},clip]{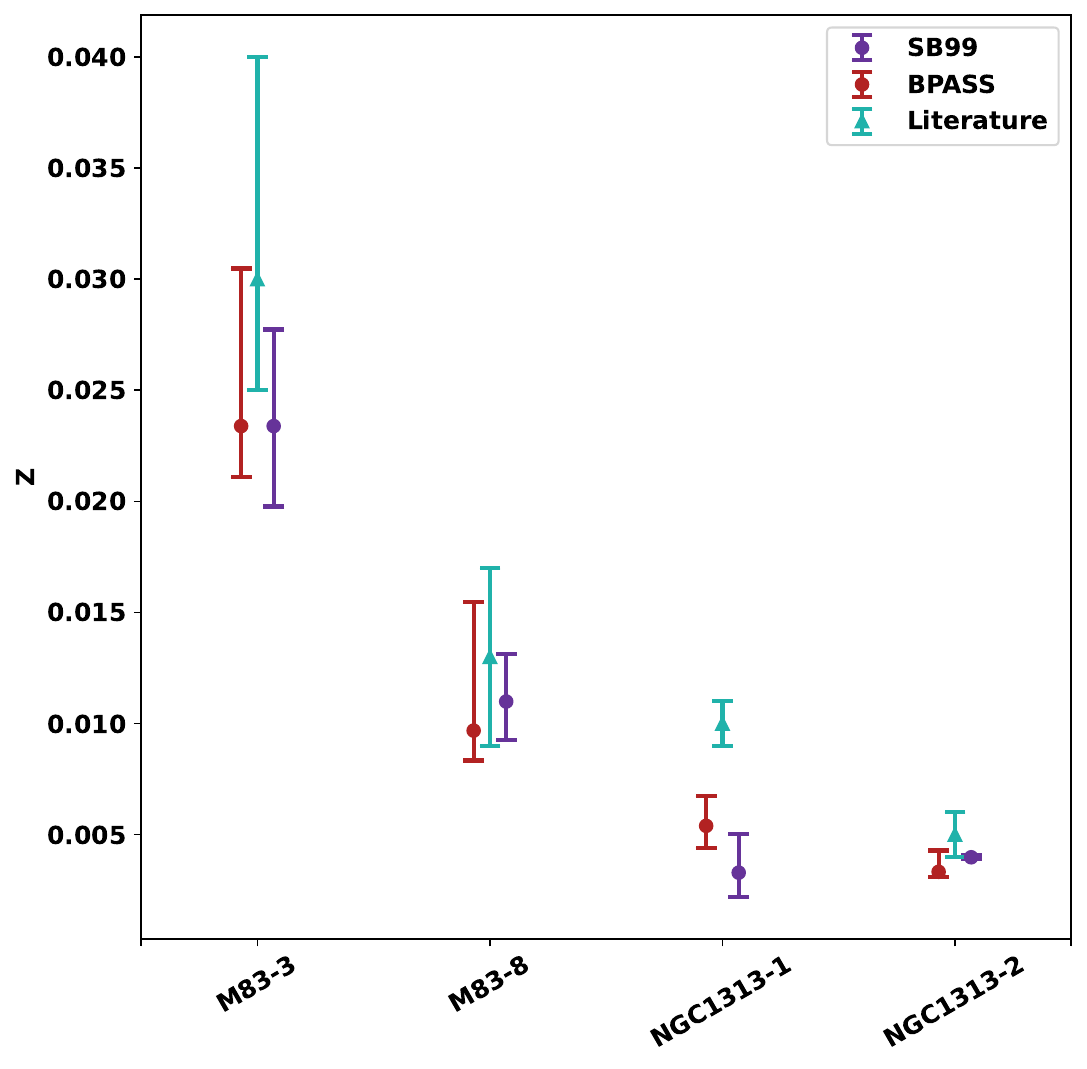}
\caption{
Comparison of the 16th, 50th, and 84th percentiles on the ages (left) and stellar metallicities (right) of the YMCs in M83 and NGC 1313 as determined by SESAMME (circles) and from other methods in the literature (teal triangles). For each cluster, we performed two SESAMME runs -- one with BPASS v2.3 models (red) and one with Starburst99 (purple). The literature values for the M83 YMCs are from \citet{Hernandez19}, while the values for the NGC 1313 YMCs are from \citet{Sirressi22}. 
}
\label{fig:litcompare}
\end{figure*}

We compare the results from SESAMME for each of the two model sets in Figure \ref{fig:litcompare}. For each cluster, the median age and metallicity inferred by SESAMME is generally consistent (within the uncertainties) across the two model sets.  We also compare the ages and metallicities from SESAMME against literature estimates for the same clusters. For the YMCs in M83, the ages are derived from {\em HST}/WFC3 photometry in \citet{Hernandez19}, while the metallicities are determined via isochrone fitting \citep{Hernandez19}. For the YMCs in NGC 1313, the literature ages and metallicities are the light-weighted averages of the two-population SSP modeling in  \citet{Sirressi22}.

For both BPASS and SB99 models, SESAMME's agreement with the literature estimates of the cluster ages is generally good within the uncertainties. The major exception to this is the cluster NGC 1313-2, where SESAMME prefers solutions with a cluster age around $\sim$6 -- 8 Myr and reddening $E(B-V) \sim$0.28, regardless of whether we use BPASS or Starburst99 SSP models. This stands in contrast to the light-weighted age of $\sim$30 Myr and slightly lower reddening of $\sim$0.2 mag determined by \citet{Sirressi22}, who used Starburst99 model SSPs for their two-population fits. Comparing their best fit for NGC 1313-2 (the purple curve in their Figure 11) and the median model inferred from our SESAMME run with BPASS (the bottom panel of Fig. \ref{fig:appendix}), we see that each model provides a reasonably good by-eye match to the observed stellar continuum. The older model preferred by \citet{Sirressi22} is also consistent with the lack of strong Si{\sc~iv} and C{\sc~iv} wind features, but does not explain the moderate N{\sc~v} emission seen in the COS spectrum. Our preferred model(s), on the other hand, are young enough to reproduce the N{\sc~v} profile, but also predict P-Cygni-like profiles for the Si{\sc~iv} and C{\sc~iv} lines, which is not observed. This may be an indication that this cluster indeed consists of two populations -- a dominant component that is $>20$ Myr old, and a small, very young component ($\sim1-3$ Myr) that drives the weak N{\sc~v} emission.


We also note that the age uncertainties when using Starburst99 are generally smaller than when using BPASS, which is due in part to the finer and linearly-uniform age grid of our Starburst99 model suite ($\Delta$age = 1 Myr) compared to that of BPASS ($\Delta$log(age) = 0.1). Finally, the inclusion of binary star physics in BPASS can cause a population to remain UV-luminous at slightly older ages compared to single-star models. This may explain why SESAMME considers older populations to be plausible solutions when using BPASS but not Starburst99 (i.e., a tail to older ages or a secondary maximum in the age posterior PDF, which leads to the asymmetric errorbars on the red points in Figure \ref{fig:litcompare}).

\section{Discussion and Summary}
\label{sec:discussion}

\begin{figure}[t]
\centering
\includegraphics[width=\linewidth,trim={0.15cm, 0.15cm, 0.15cm, 0.15cm}, clip]{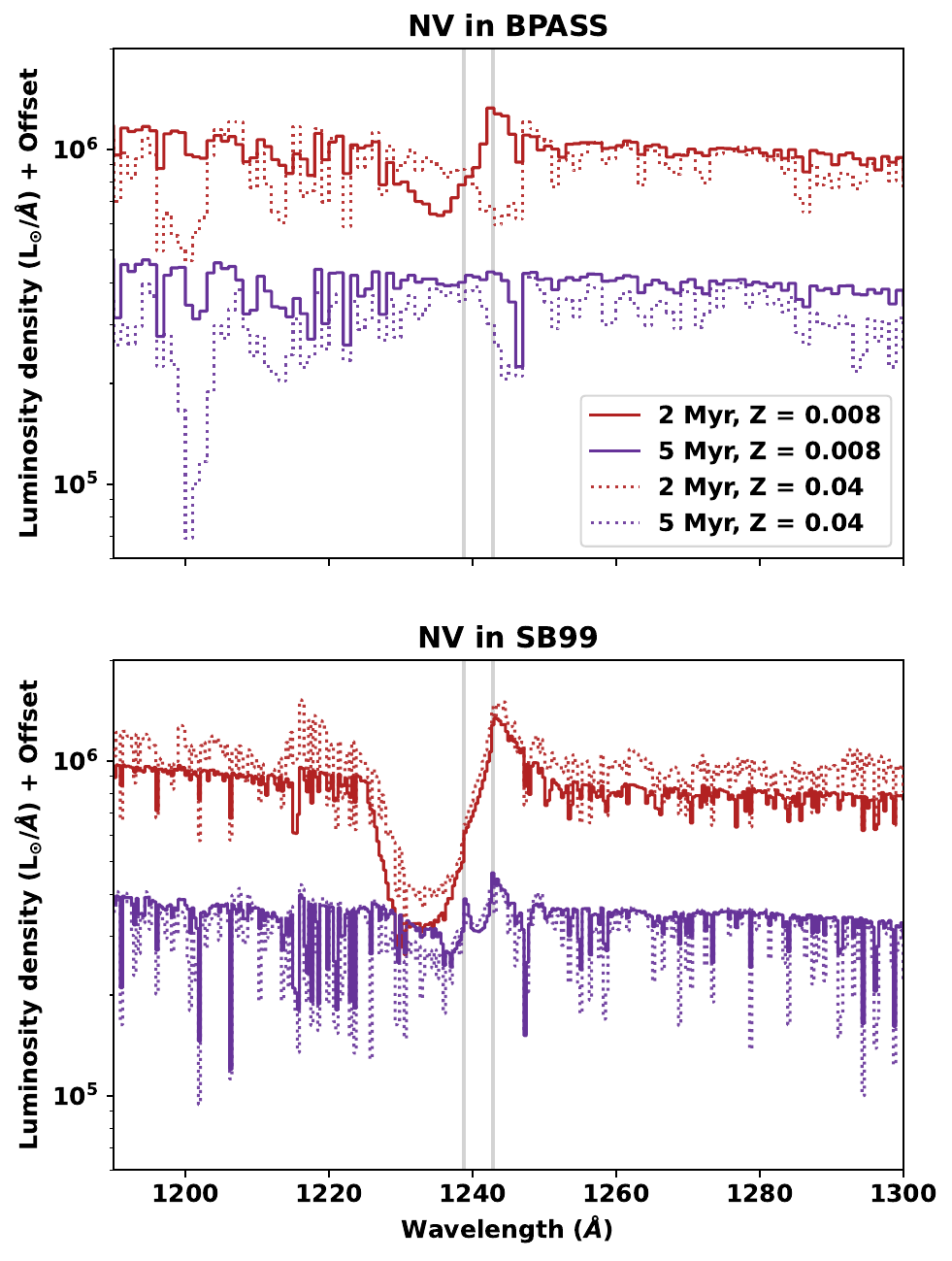}
\caption{
Comparison of N{\sc~v}$\lambda\lambda 1238,1242$ line profiles from BPASS v2.3 ({\em top}) and Starburst99 ({\em bottom}) models. Red and purple curves respectively show populations of age 2 and 5 Myr. Populations with a stellar metallicity $Z = 0.008$ are shown as solid lines, while $Z = 0.04$ models are shown with dashed lines. Some models have been shifted in luminosity for visualization purposes. The vertical grey lines mark the central wavelengths of the N{\sc~v} doublet components. Starburst99 predicts a significant P-Cygni line profile even at high metallicity and ages $\sim$5 Myr, while N{\sc~v} is seemingly absent in all but the youngest and most metal-poor BPASS models we show here.
\looseness = -2
}
\label{fig:NVexample}
\end{figure}

One of the driving principles behind the design of SESAMME is that it should be highly flexible in terms of what kinds of models it can accept as input, which in turn enables it to be useful in a greater variety of science cases. In particular, our focus on the rest-frame FUV in this work (and in a forthcoming work in which we use SESAMME to study the homogeneity of stellar and ionized-gas metallicities in nearby SFGs) is especially relevant for ongoing studies of the stellar populations of galaxies at $z > 3$, where the UV output of YMCs is shifted into optical and IR bandpasses (e.g., the VANDELS survey of \citet{Pentericci18,Cullen19} or the explosion of recent results from {\em JWST} spectroscopy). For example, unlike past versions, stellar population models in BPASS v2.3 are allowed to vary in their level of $\alpha$ enhancement at a fixed overall metallicity mass fraction $Z$ (i.e., for a fixed $Z$, increases in [$\alpha$/Fe] are offset by decreases in [Fe/H] and vice versa). This allows SESAMME to be used to constrain the  properties of stellar populations in a greater variety of environments, including clusters in starbursting galaxies and other systems that are likely analogs of higher-redshift objects.

On the other hand, stellar population models in BPASS v2.3 are only computed for a single assumed IMF, compared to the nine supplied as part of the v2.2 release. Using BPASS v2.3 may thus present issues when modeling the IL of old or intermediate-age clusters, where assumptions about the low mass end of the stellar IMF may become important in the interpretation of spectra. However, because the IMFs used in BPASS v2.2 do not vary much at the high mass end, the choice of IMF should not greatly affect the modeling of rest-frame UV spectra, which are dominated by the most massive stars. 

Because SESAMME performs a full spectral fit to constrain the age and metallicity of a cluster, it has the advantage of not being dependent on the predicted strength of any single stellar absorption or wind feature, which may be model-dependent. We illustrate the importance of this in Figure \ref{fig:NVexample}, where we show a series of models centered around the N{\sc~v}$\lambda\lambda 1238,1242$ wind feature, which is known to be present only in fairly young ($\lesssim$5 Myr) populations. The lower panel of Figure \ref{fig:NVexample} shows the N{\sc~v} profiles that are predicted by Starburst99 for a single burst of mass 10$^6 M_{\odot}$, assuming the high mass-loss Geneva stellar tracks and a double-power law IMF with the default values for the slopes and mass cutoffs. It is clear that, for a 2 Myr old population, Starburst99 predicts a stark P-Cygni profile at both low ($Z = 0.5Z_{\odot}$) and high ($Z = 2Z_{\odot}$) metallicity. Once a population reaches 5 Myr old, the absorption component of  N{\sc~v} has weakened significantly, but there are still hints of a P-Cygni profile across a range of metallicities. In the upper panel of the same figure, we show the BPASS v2.3 models with binary physics and solar abundance ratios for the same range of ages and metallicities. Unlike in Starburst99, P-Cygni-like emission for N{\sc~v} is seemingly only predicted for populations that are very young and fairly metal-poor. The lack of obvious N{\sc~v} emission from a high metallicity model population, even at ages $\sim$2 Myr, is somewhat unusual and could lead to an inaccurate characterization of a YMC if N{\sc~v} is used as the primary age diagnostic in solar or super-solar metallicity environments. (See also the median model for cluster M83-3 in Fig. \ref{fig:appendix}, which fails to reproduce the N{\sc~v} profile despite adequately tracing the continuum and other features like the Si{\sc~iv} wind line.)

We also note that all of our tests (with both mock and real FUV IL spectra) focus on relatively massive star clusters with $M_* \sim 10^4 - 10^5 M_{\odot}$, such that the IMF should be well-sampled across the full mass range. However, \citet{Orozco22} recently showed that stochastic sampling of the IMF in low-mass star clusters can drastically affect the cluster's UV light output. For example, they report that if the high-mass end of the IMF is not well-sampled, then a dearth of OB or Wolf-Rayet stars can lead to highly uncertain cluster age determinations from UBVI photometry, even in clusters as massive as $10^5 M_{\odot}$. Although their conclusions were based on photometry of massive star clusters, we expect that a similar effect may be seen while modeling the FUV spectra of low-mass YMCs. However, since the most frequently-used stellar population codes assume that the IMF is sampled deterministically, this variance in the number of very massive and UV-bright stars is difficult to account for. 

Finally, we emphasize again that SESAMME is designed to model the IL spectra of individual clusters using single SSP fitting (i.e., every star cluster assumed to be the product of a single burst of star formation). As such, it is {\em not} meant to be a spectral fitting tool for entire galaxies, for which one must account for a potentially complex star-formation history. Although SESAMME can be used to characterize YMCs in relatively distant galaxies, we caution that any such observations must have sufficient spatial resolution to isolate the YMC from its environs (to mitigate significant light contribution from, for example, an underlying old population).

In summary, we have presented the v1.0 public release of SESAMME, a flexible Python tool for Simultaneous Estimates of Star-cluster Age, Metallicity, Mass, and Extinction. In this work, we described the structure and functionality of SESAMME as a full spectral fitting tool for IL spectroscopy of star clusters, including validation tests on both synthetic (Sec. \ref{sec:validation}) and empirical (Sec. \ref{sec:cosspec}) FUV spectra of young, massive clusters. The test runs with synthetic YMC spectra, built from noise-added BPASS v2.3 stellar population models, show that SESAMME can successfully recover the known age and metallicity of the model, even at moderate signal-to-noise ratios $\sim$5 \AA$^{-1}$. Our tests with empirical FUV spectra from {\em HST}/COS, which included two sightlines towards YMCs in M83 and two in NGC 1313, were similarly successful. The ages and metallicities inferred by SESAMME for these clusters are both i) consistent with literature estimates that use independent methods, and ii) internally consistent across different assumed stellar population model sets. With the utility of SESAMME thus established in the FUV, we plan to apply it to a larger sample of extragalactic star clusters with IL spectroscopy from {\em HST}/COS (Jones et al., in prep.). SESAMME is modular and open-source, and we encourage contributions at all levels from the community.

\acknowledgements

We thank Claus Leitherer for helpful discussions regarding Starburst99 and Russell Ryan for helpful discussions on the use of \emcee{}. We also thank the anonymous referee for their careful review and suggestions, which improved the contents of this manuscript. SESAMME makes use of Astropy \citep{Astropy13,Astropy18,Astropy22}, \emcee{} \citep{emcee13}, {\tt dust\_extinction} \citep{GordonDust}, {\tt extinction} \citep{BarbaryExt}, and {\tt NumPy} \citep{Numpy11} in its computations, as well as {\tt Matplotlib} \citep{Matplotlib07} and {\tt Jupyter} \citep{JupyterNotebook16} for data visualizations. This work additionally makes use of {\tt SciPy} \citep{Scipy20} in our analyses of empirical {\em HST}/COS spectra and of v2.3 of the Binary Population and Spectral Synthesis (BPASS) models as last described in \citet{bpassv23} and \citet{bpassv22}. S.\ H.\ and B.\ L.\ J.\ acknowledge support from the European Space Agency (ESA). Support for this work was provided by NASA through grant AR-16130 from the Space Telescope Science Institute, which is operated by AURA, Inc., under NASA contract NAS5-26555. All of the data presented in this paper were obtained from MAST at the Space Telescope Science Institute. The specific observations analyzed can be accessed via~\dataset[DOI: 10.17909/mnp4-f880]{https://doi.org/10.17909/mnp4-f880}.

\bibliography{starbib,ismbib,software}

\begin{thebibliography}{}
\expandafter\ifx\csname natexlab\endcsname\relax\def\natexlab#1{#1}\fi
\providecommand{\url}[1]{\href{#1}{#1}}
\providecommand{\dodoi}[1]{doi:~\href{http://doi.org/#1}{\nolinkurl{#1}}}
\providecommand{\doeprint}[1]{\href{http://ascl.net/#1}{\nolinkurl{http://ascl.net/#1}}}
\providecommand{\doarXiv}[1]{\href{https://arxiv.org/abs/#1}{\nolinkurl{https://arxiv.org/abs/#1}}}

\bibitem[{{Aller}(1984)}]{aller84}
{Aller}, L.~H. 1984, {Physics of thermal gaseous nebulae},
  \dodoi{10.1007/978-94-010-9639-3}

\bibitem[{{Asa'd} {et~al.}(2022){Asa'd}, {Hernandez}, {As'ad}, {Molero},
  {Matteucci}, {Larsen}, \& {Chilingarian}}]{Asad22}
{Asa'd}, R., {Hernandez}, S., {As'ad}, A., {et~al.} 2022, \apj, 929, 174,
  \dodoi{10.3847/1538-4357/ac5f3e}

\bibitem[{Asplund {et~al.}(2009)Asplund, Grevesse, Sauval, \&
  Scott}]{Asplund09}
Asplund, M., Grevesse, N., Sauval, A.~J., \& Scott, P. 2009, Annual Review of
  Astronomy and Astrophysics, 47, 481,
  \dodoi{10.1146/annurev.astro.46.060407.145222}

\bibitem[{{Astropy Collaboration} {et~al.}(2013){Astropy Collaboration},
  {Robitaille}, {Tollerud}, {Greenfield}, {Droettboom}, {Bray}, {Aldcroft},
  {Davis}, {Ginsburg}, {Price-Whelan}, {Kerzendorf}, {Conley}, {Crighton},
  {Barbary}, {Muna}, {Ferguson}, {Grollier}, {Parikh}, {Nair}, {Unther},
  {Deil}, {Woillez}, {Conseil}, {Kramer}, {Turner}, {Singer}, {Fox}, {Weaver},
  {Zabalza}, {Edwards}, {Azalee Bostroem}, {Burke}, {Casey}, {Crawford},
  {Dencheva}, {Ely}, {Jenness}, {Labrie}, {Lim}, {Pierfederici}, {Pontzen},
  {Ptak}, {Refsdal}, {Servillat}, \& {Streicher}}]{Astropy13}
{Astropy Collaboration}, {Robitaille}, T.~P., {Tollerud}, E.~J., {et~al.} 2013,
  \aap, 558, A33, \dodoi{10.1051/0004-6361/201322068}

\bibitem[{{Astropy Collaboration} {et~al.}(2018){Astropy Collaboration},
  {Price-Whelan}, {Sip{\H{o}}cz}, {G{\"u}nther}, {Lim}, {Crawford}, {Conseil},
  {Shupe}, {Craig}, {Dencheva}, {Ginsburg}, {VanderPlas}, {Bradley},
  {P{\'e}rez-Su{\'a}rez}, {de Val-Borro}, {Aldcroft}, {Cruz}, {Robitaille},
  {Tollerud}, {Ardelean}, {Babej}, {Bach}, {Bachetti}, {Bakanov}, {Bamford},
  {Barentsen}, {Barmby}, {Baumbach}, {Berry}, {Biscani}, {Boquien}, {Bostroem},
  {Bouma}, {Brammer}, {Bray}, {Breytenbach}, {Buddelmeijer}, {Burke},
  {Calderone}, {Cano Rodr{\'\i}guez}, {Cara}, {Cardoso}, {Cheedella}, {Copin},
  {Corrales}, {Crichton}, {D'Avella}, {Deil}, {Depagne}, {Dietrich}, {Donath},
  {Droettboom}, {Earl}, {Erben}, {Fabbro}, {Ferreira}, {Finethy}, {Fox},
  {Garrison}, {Gibbons}, {Goldstein}, {Gommers}, {Greco}, {Greenfield},
  {Groener}, {Grollier}, {Hagen}, {Hirst}, {Homeier}, {Horton}, {Hosseinzadeh},
  {Hu}, {Hunkeler}, {Ivezi{\'c}}, {Jain}, {Jenness}, {Kanarek}, {Kendrew},
  {Kern}, {Kerzendorf}, {Khvalko}, {King}, {Kirkby}, {Kulkarni}, {Kumar},
  {Lee}, {Lenz}, {Littlefair}, {Ma}, {Macleod}, {Mastropietro}, {McCully},
  {Montagnac}, {Morris}, {Mueller}, {Mumford}, {Muna}, {Murphy}, {Nelson},
  {Nguyen}, {Ninan}, {N{\"o}the}, {Ogaz}, {Oh}, {Parejko}, {Parley}, {Pascual},
  {Patil}, {Patil}, {Plunkett}, {Prochaska}, {Rastogi}, {Reddy Janga},
  {Sabater}, {Sakurikar}, {Seifert}, {Sherbert}, {Sherwood-Taylor}, {Shih},
  {Sick}, {Silbiger}, {Singanamalla}, {Singer}, {Sladen}, {Sooley},
  {Sornarajah}, {Streicher}, {Teuben}, {Thomas}, {Tremblay}, {Turner},
  {Terr{\'o}n}, {van Kerkwijk}, {de la Vega}, {Watkins}, {Weaver}, {Whitmore},
  {Woillez}, {Zabalza}, \& {Astropy Contributors}}]{Astropy18}
{Astropy Collaboration}, {Price-Whelan}, A.~M., {Sip{\H{o}}cz}, B.~M., {et~al.}
  2018, \aj, 156, 123, \dodoi{10.3847/1538-3881/aabc4f}

\bibitem[{{Astropy Collaboration} {et~al.}(2022){Astropy Collaboration},
  {Price-Whelan}, {Lim}, {Earl}, {Starkman}, {Bradley}, {Shupe}, {Patil},
  {Corrales}, {Brasseur}, {N{\"o}the}, {Donath}, {Tollerud}, {Morris},
  {Ginsburg}, {Vaher}, {Weaver}, {Tocknell}, {Jamieson}, {van Kerkwijk},
  {Robitaille}, {Merry}, {Bachetti}, {G{\"u}nther}, {Aldcroft},
  {Alvarado-Montes}, {Archibald}, {B{\'o}di}, {Bapat}, {Barentsen},
  {Baz{\'a}n}, {Biswas}, {Boquien}, {Burke}, {Cara}, {Cara}, {Conroy},
  {Conseil}, {Craig}, {Cross}, {Cruz}, {D'Eugenio}, {Dencheva}, {Devillepoix},
  {Dietrich}, {Eigenbrot}, {Erben}, {Ferreira}, {Foreman-Mackey}, {Fox},
  {Freij}, {Garg}, {Geda}, {Glattly}, {Gondhalekar}, {Gordon}, {Grant},
  {Greenfield}, {Groener}, {Guest}, {Gurovich}, {Handberg}, {Hart},
  {Hatfield-Dodds}, {Homeier}, {Hosseinzadeh}, {Jenness}, {Jones}, {Joseph},
  {Kalmbach}, {Karamehmetoglu}, {Ka{\l}uszy{\'n}ski}, {Kelley}, {Kern},
  {Kerzendorf}, {Koch}, {Kulumani}, {Lee}, {Ly}, {Ma}, {MacBride}, {Maljaars},
  {Muna}, {Murphy}, {Norman}, {O'Steen}, {Oman}, {Pacifici}, {Pascual},
  {Pascual-Granado}, {Patil}, {Perren}, {Pickering}, {Rastogi}, {Roulston},
  {Ryan}, {Rykoff}, {Sabater}, {Sakurikar}, {Salgado}, {Sanghi}, {Saunders},
  {Savchenko}, {Schwardt}, {Seifert-Eckert}, {Shih}, {Jain}, {Shukla}, {Sick},
  {Simpson}, {Singanamalla}, {Singer}, {Singhal}, {Sinha}, {Sip{\H{o}}cz},
  {Spitler}, {Stansby}, {Streicher}, {{\v{S}}umak}, {Swinbank}, {Taranu},
  {Tewary}, {Tremblay}, {de Val-Borro}, {Van Kooten}, {Vasovi{\'c}}, {Verma},
  {de Miranda Cardoso}, {Williams}, {Wilson}, {Winkel}, {Wood-Vasey}, {Xue},
  {Yoachim}, {Zhang}, {Zonca}, \& {Astropy Project Contributors}}]{Astropy22}
{Astropy Collaboration}, {Price-Whelan}, A.~M., {Lim}, P.~L., {et~al.} 2022,
  \apj, 935, 167, \dodoi{10.3847/1538-4357/ac7c74}

\bibitem[{{Barbary}(2021)}]{BarbaryExt}
{Barbary}, K. 2021, {extinction: Dust extinction laws}, Astrophysics Source
  Code Library, record ascl:2102.026.
\newblock \doeprint{2102.026}

\bibitem[{{Bresolin} {et~al.}(2016){Bresolin}, {Kudritzki}, {Urbaneja},
  {Gieren}, {Ho}, \& {Pietrzy{\'n}ski}}]{Bresolin16}
{Bresolin}, F., {Kudritzki}, R.-P., {Urbaneja}, M.~A., {et~al.} 2016, \apj,
  830, 64, \dodoi{10.3847/0004-637X/830/2/64}

\bibitem[{{Byrne} {et~al.}(2022){Byrne}, {Stanway}, {Eldridge}, {McSwiney}, \&
  {Townsend}}]{bpassv23}
{Byrne}, C.~M., {Stanway}, E.~R., {Eldridge}, J.~J., {McSwiney}, L., \&
  {Townsend}, O.~T. 2022, \mnras, 512, 5329, \dodoi{10.1093/mnras/stac807}

\bibitem[{{Calzetti} {et~al.}(2000){Calzetti}, {Armus}, {Bohlin}, {Kinney},
  {Koornneef}, \& {Storchi-Bergmann}}]{calzetti2000}
{Calzetti}, D., {Armus}, L., {Bohlin}, R.~C., {et~al.} 2000, \apj, 533, 682,
  \dodoi{10.1086/308692}

\bibitem[{{Cardelli} {et~al.}(1989){Cardelli}, {Clayton}, \& {Mathis}}]{ccm89}
{Cardelli}, J.~A., {Clayton}, G.~C., \& {Mathis}, J.~S. 1989, \apj, 345, 245,
  \dodoi{10.1086/167900}

\bibitem[{{Chabrier}(2003)}]{chab03}
{Chabrier}, G. 2003, \pasp, 115, 763, \dodoi{10.1086/376392}

\bibitem[{{Chisholm} {et~al.}(2019){Chisholm}, {Rigby}, {Bayliss}, {Berg},
  {Dahle}, {Gladders}, \& {Sharon}}]{chisholm19}
{Chisholm}, J., {Rigby}, J.~R., {Bayliss}, M., {et~al.} 2019, \apj, 882, 182,
  \dodoi{10.3847/1538-4357/ab3104}

\bibitem[{{Colucci} {et~al.}(2011){Colucci}, {Bernstein}, {Cameron}, \&
  {McWilliam}}]{Colucci11}
{Colucci}, J.~E., {Bernstein}, R.~A., {Cameron}, S.~A., \& {McWilliam}, A.
  2011, \apj, 735, 55, \dodoi{10.1088/0004-637X/735/1/55}

\bibitem[{{Colucci} {et~al.}(2012){Colucci}, {Bernstein}, {Cameron}, \&
  {McWilliam}}]{Colucci12}
---. 2012, \apj, 746, 29, \dodoi{10.1088/0004-637X/746/1/29}

\bibitem[{{Cullen} {et~al.}(2019){Cullen}, {McLure}, {Dunlop}, {Khochfar},
  {Dav{\'e}}, {Amor{\'\i}n}, {Bolzonella}, {Carnall}, {Castellano}, {Cimatti},
  {Cirasuolo}, {Cresci}, {Fynbo}, {Fontanot}, {Gargiulo}, {Garilli}, {Guaita},
  {Hathi}, {Hibon}, {Mannucci}, {Marchi}, {McLeod}, {Pentericci}, {Pozzetti},
  {Shapley}, {Talia}, \& {Zamorani}}]{Cullen19}
{Cullen}, F., {McLure}, R.~J., {Dunlop}, J.~S., {et~al.} 2019, \mnras, 487,
  2038, \dodoi{10.1093/mnras/stz1402}

\bibitem[{{Danforth} {et~al.}(2010){Danforth}, {Keeney}, {Stocke}, {Shull}, \&
  {Yao}}]{DanforthCOS}
{Danforth}, C.~W., {Keeney}, B.~A., {Stocke}, J.~T., {Shull}, J.~M., \& {Yao},
  Y. 2010, \apj, 720, 976, \dodoi{10.1088/0004-637X/720/1/976}

\bibitem[{{Davies} {et~al.}(2010){Davies}, {Kudritzki}, \& {Figer}}]{Davies10}
{Davies}, B., {Kudritzki}, R.-P., \& {Figer}, D.~F. 2010, \mnras, 407, 1203,
  \dodoi{10.1111/j.1365-2966.2010.16965.x}

\bibitem[{{Davies} {et~al.}(2015){Davies}, {Kudritzki}, {Gazak}, {Plez},
  {Bergemann}, {Evans}, \& {Patrick}}]{Davies15}
{Davies}, B., {Kudritzki}, R.-P., {Gazak}, Z., {et~al.} 2015, \apj, 806, 21,
  \dodoi{10.1088/0004-637X/806/1/21}

\bibitem[{{Davies} {et~al.}(2017){Davies}, {Kudritzki}, {Lardo}, {Bergemann},
  {Beasor}, {Plez}, {Evans}, {Bastian}, \& {Patrick}}]{Davies17}
{Davies}, B., {Kudritzki}, R.-P., {Lardo}, C., {et~al.} 2017, \apj, 847, 112,
  \dodoi{10.3847/1538-4357/aa89ed}

\bibitem[{{Ferland}(1980)}]{ferland80}
{Ferland}, G.~J. 1980, \pasp, 92, 596, \dodoi{10.1086/130718}

\bibitem[{{Fitzpatrick}(1999)}]{fitzpatrick99}
{Fitzpatrick}, E.~L. 1999, \pasp, 111, 63, \dodoi{10.1086/316293}

\bibitem[{{Fitzpatrick} \& {Massa}(2007)}]{fm07}
{Fitzpatrick}, E.~L., \& {Massa}, D. 2007, \apj, 663, 320,
  \dodoi{10.1086/518158}

\bibitem[{{Foreman-Mackey} {et~al.}(2013){Foreman-Mackey}, {Hogg}, {Lang}, \&
  {Goodman}}]{emcee13}
{Foreman-Mackey}, D., {Hogg}, D.~W., {Lang}, D., \& {Goodman}, J. 2013, \pasp,
  125, 306, \dodoi{10.1086/670067}

\bibitem[{{Gazak} {et~al.}(2014){Gazak}, {Davies}, {Bastian}, {Kudritzki},
  {Bergemann}, {Plez}, {Evans}, {Patrick}, {Bresolin}, \&
  {Schinnerer}}]{Gazak14}
{Gazak}, J.~Z., {Davies}, B., {Bastian}, N., {et~al.} 2014, \apj, 787, 142,
  \dodoi{10.1088/0004-637X/787/2/142}

\bibitem[{Gordon {et~al.}(2022)Gordon, Larson, McBride, Lim, Sipőcz, \&
  Gaikwad}]{GordonDust}
Gordon, K., Larson, K., McBride, A., {et~al.} 2022, {karllark/dust\_extinction:
  NIRSpectralExtinctionAdded}, v1.1,  Zenodo, \dodoi{10.5281/zenodo.6397654}

\bibitem[{{Gordon} {et~al.}(2023){Gordon}, {Clayton}, {Decleir}, {Fitzpatrick},
  {Massa}, {Misselt}, \& {Tollerud}}]{Gordon23}
{Gordon}, K.~D., {Clayton}, G.~C., {Decleir}, M., {et~al.} 2023, \apj, 950, 86,
  \dodoi{10.3847/1538-4357/accb59}

\bibitem[{{Gordon} {et~al.}(2003){Gordon}, {Clayton}, {Misselt}, {Landolt}, \&
  {Wolff}}]{gordon03}
{Gordon}, K.~D., {Clayton}, G.~C., {Misselt}, K.~A., {Landolt}, A.~U., \&
  {Wolff}, M.~J. 2003, \apj, 594, 279, \dodoi{10.1086/376774}

\bibitem[{{Grevesse} \& {Noels}(1993)}]{Grevesse93}
{Grevesse}, N., \& {Noels}, A. 1993, in Origin and Evolution of the Elements,
  ed. N.~{Prantzos}, E.~{Vangioni-Flam}, \& M.~{Casse}, 15--25

\bibitem[{{Gvozdenko} {et~al.}(2022){Gvozdenko}, {Larsen}, {Beasley}, \&
  {Brodie}}]{Gvozdenko22}
{Gvozdenko}, A., {Larsen}, S.~S., {Beasley}, M.~A., \& {Brodie}, J. 2022, \aap,
  666, A159, \dodoi{10.1051/0004-6361/202243415}

\bibitem[{{Halliday} {et~al.}(2008){Halliday}, {Daddi}, {Cimatti}, {Kurk},
  {Renzini}, {Mignoli}, {Bolzonella}, {Pozzetti}, {Dickinson}, {Zamorani},
  {Berta}, {Franceschini}, {Cassata}, {Rodighiero}, \& {Rosati}}]{Halliday08}
{Halliday}, C., {Daddi}, E., {Cimatti}, A., {et~al.} 2008, \aap, 479, 417,
  \dodoi{10.1051/0004-6361:20078673}

\bibitem[{{Hernandez} {et~al.}(2020){Hernandez}, {Aloisi}, {James}, {Ferland},
  {Fox}, {Tosi}, \& {Tumlinson}}]{Hernandez20}
{Hernandez}, S., {Aloisi}, A., {James}, B.~L., {et~al.} 2020, \apj, 892, 19,
  \dodoi{10.3847/1538-4357/ab77c6}

\bibitem[{{Hernandez} {et~al.}(2017){Hernandez}, {Larsen}, {Trager}, {Groot},
  \& {Kaper}}]{Hernandez17}
{Hernandez}, S., {Larsen}, S., {Trager}, S., {Groot}, P., \& {Kaper}, L. 2017,
  \aap, 603, A119, \dodoi{10.1051/0004-6361/201730550}

\bibitem[{{Hernandez} {et~al.}(2022){Hernandez}, {Winch}, {Larsen}, {James}, \&
  {Jones}}]{Hernandez22}
{Hernandez}, S., {Winch}, A., {Larsen}, S., {James}, B.~L., \& {Jones}, L.
  2022, \aj, 164, 89, \dodoi{10.3847/1538-3881/ac7ebe}

\bibitem[{{Hernandez} {et~al.}(2019){Hernandez}, {Larsen}, {Aloisi}, {Berg},
  {Blair}, {Fox}, {Heckman}, {James}, {Long}, {Skillman}, \&
  {Whitmore}}]{Hernandez19}
{Hernandez}, S., {Larsen}, S., {Aloisi}, A., {et~al.} 2019, \apj, 872, 116,
  \dodoi{10.3847/1538-4357/ab017a}

\bibitem[{{Hernandez} {et~al.}(2021){Hernandez}, {Aloisi}, {James}, {Kumari},
  {Berg}, {Adamo}, {Blair}, {Faucher-Gigu{\`e}re}, {Fox}, {Gurvich}, {Hafen},
  {Heckman}, {Lebouteiller}, {Long}, {Skillman}, {Tumlinson}, \&
  {Whitmore}}]{Hernandez21}
{Hernandez}, S., {Aloisi}, A., {James}, B.~L., {et~al.} 2021, \apj, 908, 226,
  \dodoi{10.3847/1538-4357/abd6c4}

\bibitem[{{Ho} {et~al.}(2015){Ho}, {Kudritzki}, {Kewley}, {Zahid}, {Dopita},
  {Bresolin}, \& {Rupke}}]{Ho15}
{Ho}, I.~T., {Kudritzki}, R.-P., {Kewley}, L.~J., {et~al.} 2015, \mnras, 448,
  2030, \dodoi{10.1093/mnras/stv067}

\bibitem[{{Hosek} {et~al.}(2014){Hosek}, {Kudritzki}, {Bresolin}, {Urbaneja},
  {Evans}, {Pietrzy{\'n}ski}, {Gieren}, {Przybilla}, \& {Carraro}}]{Hosek14}
{Hosek}, Matthew~W., J., {Kudritzki}, R.-P., {Bresolin}, F., {et~al.} 2014,
  \apj, 785, 151, \dodoi{10.1088/0004-637X/785/2/151}

\bibitem[{{Hunter}(2007)}]{Matplotlib07}
{Hunter}, J.~D. 2007, Computing in Science and Engineering, 9, 90,
  \dodoi{10.1109/MCSE.2007.55}

\bibitem[{{James} {et~al.}(2014){James}, {Aloisi}, {Heckman}, {Sohn}, \&
  {Wolfe}}]{James14}
{James}, B.~L., {Aloisi}, A., {Heckman}, T., {Sohn}, S.~T., \& {Wolfe}, M.~A.
  2014, \apj, 795, 109, \dodoi{10.1088/0004-637X/795/2/109}

\bibitem[{{Kluyver} {et~al.}(2016){Kluyver}, {Ragan-Kelley}, {P{\'e}rez},
  {Granger}, {Bussonnier}, {Frederic}, {Kelley}, {Hamrick}, {Grout}, {Corlay},
  {Ivanov}, {Avila}, {Abdalla}, {Willing}, \& {Jupyter Development
  Team}}]{JupyterNotebook16}
{Kluyver}, T., {Ragan-Kelley}, B., {P{\'e}rez}, F., {et~al.} 2016, in IOS
  Press, 87--90, \dodoi{10.3233/978-1-61499-649-1-87}

\bibitem[{{Krogager}(2018)}]{Voigtfit18}
{Krogager}, J.-K. 2018, {VoigtFit: Absorption line fitting for Voigt profiles},
  Astrophysics Source Code Library, record ascl:1811.016.
\newblock \doeprint{1811.016}

\bibitem[{{Kudritzki} {et~al.}(2015){Kudritzki}, {Ho}, {Schruba}, {Burkert},
  {Zahid}, {Bresolin}, \& {Dima}}]{Kudritzki15}
{Kudritzki}, R.-P., {Ho}, I.~T., {Schruba}, A., {et~al.} 2015, \mnras, 450,
  342, \dodoi{10.1093/mnras/stv522}

\bibitem[{{Kudritzki} {et~al.}(2014){Kudritzki}, {Urbaneja}, {Bresolin},
  {Hosek}, \& {Przybilla}}]{Kudritzki14}
{Kudritzki}, R.-P., {Urbaneja}, M.~A., {Bresolin}, F., {Hosek}, Matthew~W., J.,
  \& {Przybilla}, N. 2014, \apj, 788, 56, \dodoi{10.1088/0004-637X/788/1/56}

\bibitem[{{Kudritzki} {et~al.}(2012){Kudritzki}, {Urbaneja}, {Gazak},
  {Bresolin}, {Przybilla}, {Gieren}, \& {Pietrzy{\'n}ski}}]{Kudritzki12}
{Kudritzki}, R.-P., {Urbaneja}, M.~A., {Gazak}, Z., {et~al.} 2012, \apj, 747,
  15, \dodoi{10.1088/0004-637X/747/1/15}

\bibitem[{{Kunth} \& {Sargent}(1986)}]{Kunth86}
{Kunth}, D., \& {Sargent}, W.~L.~W. 1986, \apj, 300, 496,
  \dodoi{10.1086/163828}

\bibitem[{{Larsen} {et~al.}(2012){Larsen}, {Brodie}, \& {Strader}}]{Larsen12}
{Larsen}, S.~S., {Brodie}, J.~P., \& {Strader}, J. 2012, \aap, 546, A53,
  \dodoi{10.1051/0004-6361/201219895}

\bibitem[{{Larsen} {et~al.}(2018){Larsen}, {Pugliese}, \& {Brodie}}]{Larsen18b}
{Larsen}, S.~S., {Pugliese}, G., \& {Brodie}, J.~P. 2018, \aap, 617, A119,
  \dodoi{10.1051/0004-6361/201832767}

\bibitem[{{Lebouteiller} {et~al.}(2013){Lebouteiller}, {Heap}, {Hubeny}, \&
  {Kunth}}]{Lebouteiller13}
{Lebouteiller}, V., {Heap}, S., {Hubeny}, I., \& {Kunth}, D. 2013, \aap, 553,
  A16, \dodoi{10.1051/0004-6361/201220948}

\bibitem[{{Leitherer} {et~al.}(1999){Leitherer}, {Schaerer}, {Goldader},
  {Delgado}, {Robert}, {Kune}, {de Mello}, {Devost}, \&
  {Heckman}}]{starburst99}
{Leitherer}, C., {Schaerer}, D., {Goldader}, J.~D., {et~al.} 1999, \apjs, 123,
  3, \dodoi{10.1086/313233}

\bibitem[{{Maiolino} \& {Mannucci}(2019)}]{Maiolino19}
{Maiolino}, R., \& {Mannucci}, F. 2019, \aapr, 27, 3,
  \dodoi{10.1007/s00159-018-0112-2}

\bibitem[{{McWilliam} \& {Bernstein}(2008)}]{Mcwilliam08}
{McWilliam}, A., \& {Bernstein}, R.~A. 2008, \apj, 684, 326,
  \dodoi{10.1086/589957}

\bibitem[{{Orozco-Duarte} {et~al.}(2022){Orozco-Duarte}, {Wofford},
  {Vidal-Garc{\'\i}a}, {Bruzual}, {Charlot}, {Krumholz}, {Hannon}, {Lee},
  {Wofford}, {Fumagalli}, {Dale}, {Messa}, {Grebel}, {Smith}, {Grasha}, \&
  {Cook}}]{Orozco22}
{Orozco-Duarte}, R., {Wofford}, A., {Vidal-Garc{\'\i}a}, A., {et~al.} 2022,
  \mnras, 509, 522, \dodoi{10.1093/mnras/stab2988}

\bibitem[{{Pentericci} {et~al.}(2018){Pentericci}, {McLure}, {Garilli},
  {Cucciati}, {Franzetti}, {Iovino}, {Amorin}, {Bolzonella}, {Bongiorno},
  {Carnall}, {Castellano}, {Cimatti}, {Cirasuolo}, {Cullen}, {De Barros},
  {Dunlop}, {Elbaz}, {Finkelstein}, {Fontana}, {Fontanot}, {Fumana},
  {Gargiulo}, {Guaita}, {Hartley}, {Jarvis}, {Juneau}, {Karman}, {Maccagni},
  {Marchi}, {Marmol-Queralto}, {Nandra}, {Pompei}, {Pozzetti}, {Scodeggio},
  {Sommariva}, {Talia}, {Almaini}, {Balestra}, {Bardelli}, {Bell}, {Bourne},
  {Bowler}, {Brusa}, {Buitrago}, {Caputi}, {Cassata}, {Charlot}, {Citro},
  {Cresci}, {Cristiani}, {Curtis-Lake}, {Dickinson}, {Fazio}, {Ferguson},
  {Fiore}, {Franco}, {Fynbo}, {Galametz}, {Georgakakis}, {Giavalisco},
  {Grazian}, {Hathi}, {Jung}, {Kim}, {Koekemoer}, {Khusanova}, {Le F{\`e}vre},
  {Lotz}, {Mannucci}, {Maltby}, {Matsuoka}, {McLeod}, {Mendez-Hernandez},
  {Mendez-Abreu}, {Mignoli}, {Moresco}, {Mortlock}, {Nonino}, {Pannella},
  {Papovich}, {Popesso}, {Rosario}, {Salvato}, {Santini}, {Schaerer},
  {Schreiber}, {Stark}, {Tasca}, {Thomas}, {Treu}, {Vanzella}, {Wild},
  {Williams}, {Zamorani}, \& {Zucca}}]{Pentericci18}
{Pentericci}, L., {McLure}, R.~J., {Garilli}, B., {et~al.} 2018, \aap, 616,
  A174, \dodoi{10.1051/0004-6361/201833047}

\bibitem[{{Schlafly} \& {Finkbeiner}(2011)}]{Schlafly11}
{Schlafly}, E.~F., \& {Finkbeiner}, D.~P. 2011, \apj, 737, 103,
  \dodoi{10.1088/0004-637X/737/2/103}

\bibitem[{{Sirressi} {et~al.}(2022){Sirressi}, {Adamo}, {Hayes}, {Bik},
  {Strand{\"a}nger}, {Runnholm}, {Oey}, {{\"O}stlin}, {Menacho}, \&
  {Smith}}]{Sirressi22}
{Sirressi}, M., {Adamo}, A., {Hayes}, M., {et~al.} 2022, \mnras, 510, 4819,
  \dodoi{10.1093/mnras/stab3774}

\bibitem[{{Stanway} \& {Eldridge}(2018)}]{bpassv22}
{Stanway}, E.~R., \& {Eldridge}, J.~J. 2018, \mnras, 479, 75,
  \dodoi{10.1093/mnras/sty1353}

\bibitem[{{Steidel} {et~al.}(2016){Steidel}, {Strom}, {Pettini}, {Rudie},
  {Reddy}, \& {Trainor}}]{Steidel16}
{Steidel}, C.~C., {Strom}, A.~L., {Pettini}, M., {et~al.} 2016, \apj, 826, 159,
  \dodoi{10.3847/0004-637X/826/2/159}

\bibitem[{{Tremonti} {et~al.}(2004){Tremonti}, {Heckman}, {Kauffmann},
  {Brinchmann}, {Charlot}, {White}, {Seibert}, {Peng}, {Schlegel}, {Uomoto},
  {Fukugita}, \& {Brinkmann}}]{Tremonti04}
{Tremonti}, C.~A., {Heckman}, T.~M., {Kauffmann}, G., {et~al.} 2004, \apj, 613,
  898, \dodoi{10.1086/423264}

\bibitem[{{van der Walt} {et~al.}(2011){van der Walt}, {Colbert}, \&
  {Varoquaux}}]{Numpy11}
{van der Walt}, S., {Colbert}, S.~C., \& {Varoquaux}, G. 2011, Computing in
  Science and Engineering, 13, 22, \dodoi{10.1109/MCSE.2011.37}

\bibitem[{{Virtanen} {et~al.}(2020){Virtanen}, {Gommers}, {Oliphant},
  {Haberland}, {Reddy}, {Cournapeau}, {Burovski}, {Peterson}, {Weckesser},
  {Bright}, {van der Walt}, {Brett}, {Wilson}, {Millman}, {Mayorov}, {Nelson},
  {Jones}, {Kern}, {Larson}, {Carey}, {Polat}, {Feng}, {Moore}, {VanderPlas},
  {Laxalde}, {Perktold}, {Cimrman}, {Henriksen}, {Quintero}, {Harris},
  {Archibald}, {Ribeiro}, {Pedregosa}, {van Mulbregt}, \& {SciPy 1. 0
  Contributors}}]{Scipy20}
{Virtanen}, P., {Gommers}, R., {Oliphant}, T.~E., {et~al.} 2020, Nature
  Methods, 17, 261, \dodoi{10.1038/s41592-019-0686-2}

\bibitem[{{Walsh} \& {Roy}(1997)}]{walsh97}
{Walsh}, J.~R., \& {Roy}, J.~R. 1997, \mnras, 288, 726,
  \dodoi{10.1093/mnras/288.3.726}

\bibitem[{{Zaritsky} {et~al.}(1994){Zaritsky}, {Kennicutt}, \&
  {Huchra}}]{Zaritsky94}
{Zaritsky}, D., {Kennicutt}, Robert~C., J., \& {Huchra}, J.~P. 1994, \apj, 420,
  87, \dodoi{10.1086/173544}

\end{thebibliography}

\appendix
\counterwithin{figure}{section}

\section{SESAMME Fits to Remaining Clusters}

\begin{figure*}[t]
\centering
\includegraphics[width=\linewidth,height=0.52\linewidth,trim={0.1cm 0.2cm 0.1cm 0.2cm},clip]{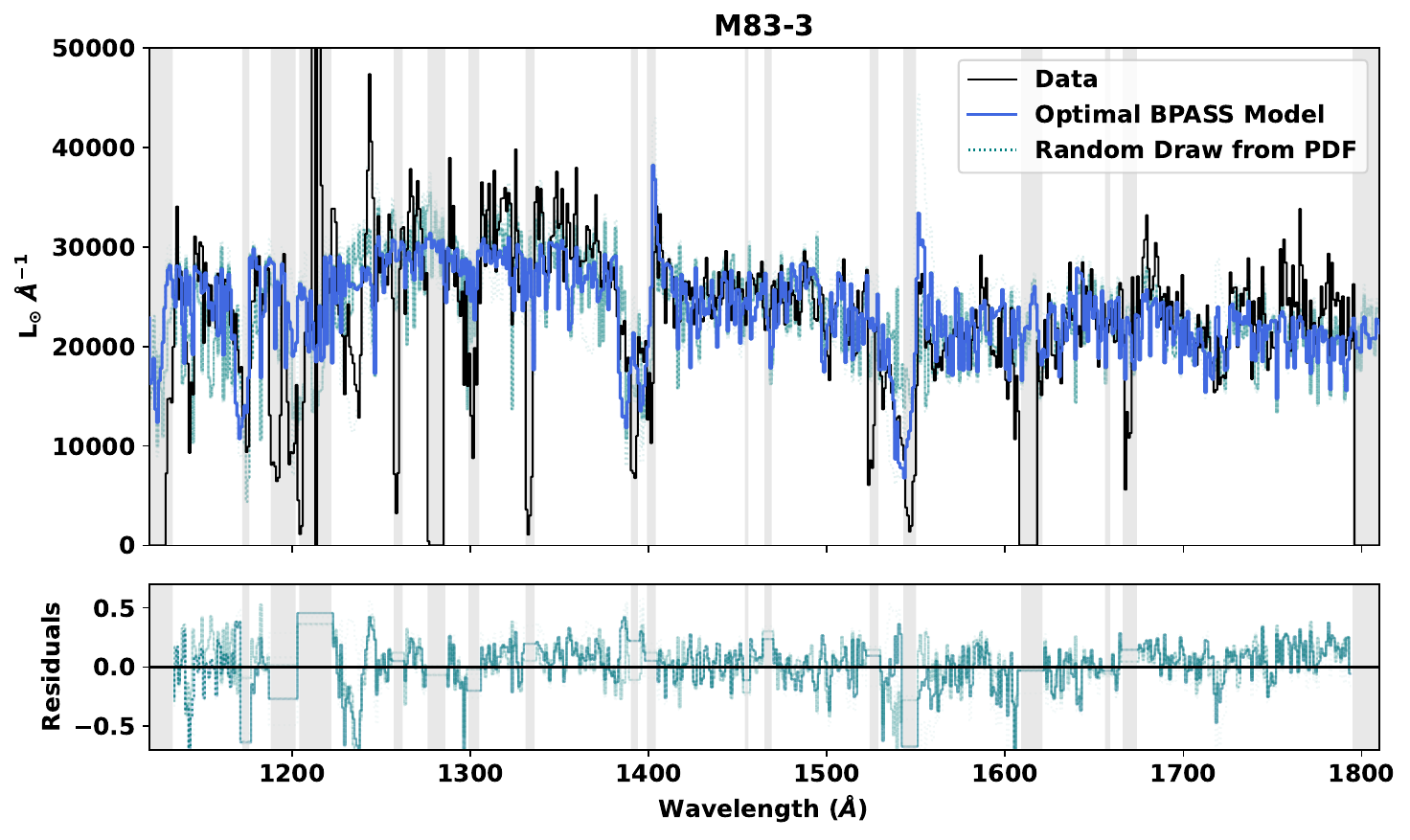}\\
\medskip
\includegraphics[width=\linewidth,height=0.52\linewidth,trim={0.1cm 0.2cm 0.1cm 0.7cm},clip]{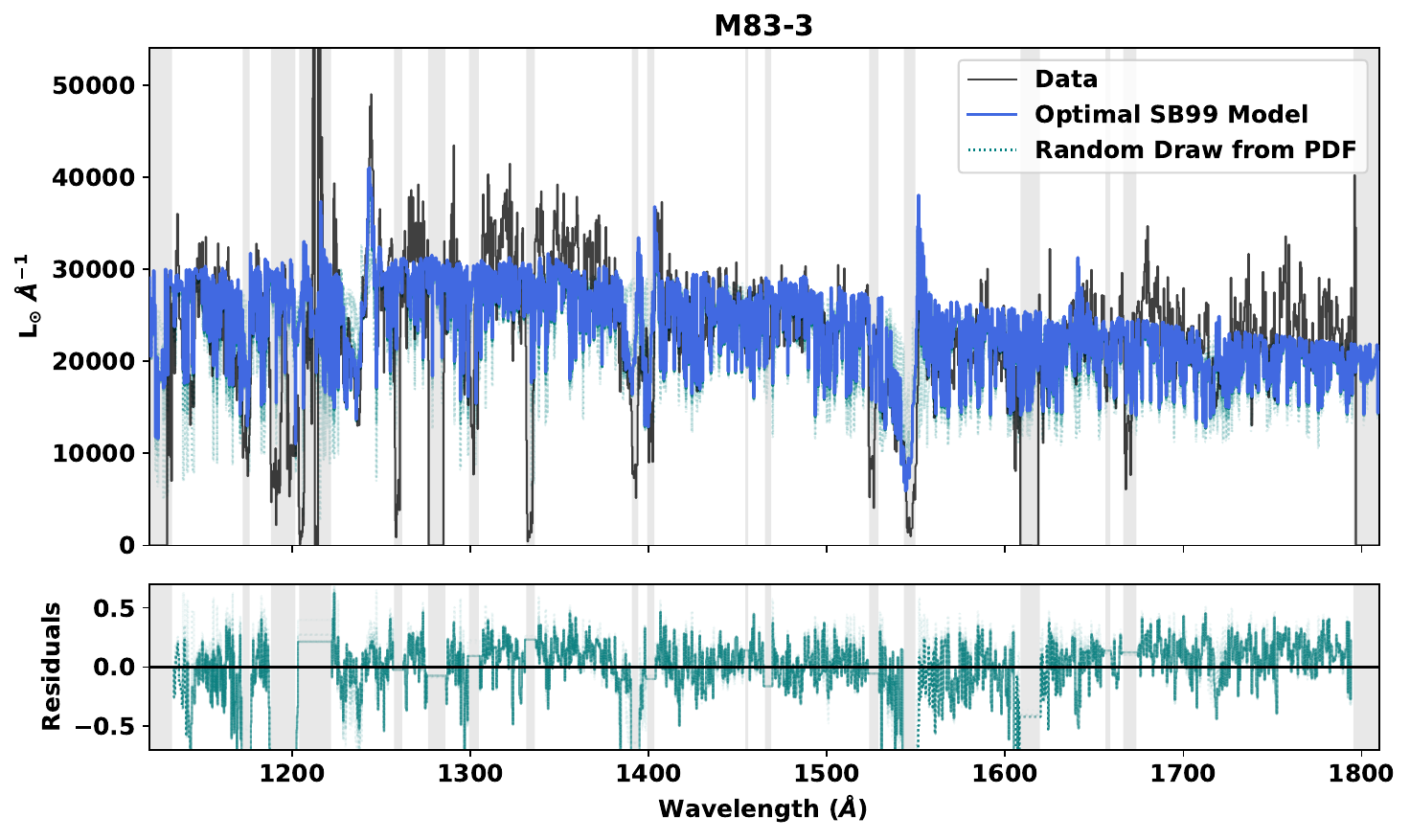}\\
\caption{Continuation of Fig. \ref{fig:M83fit} showing the performance of SESAMME in fitting FUV spectra of YMCs. {\em Top:} The black curve shows the H{\sc~i} absorption-corrected, combined G130M+G160M spectrum of YMC M83-3 \citep{Hernandez19} from {\em HST}/COS, after smoothing and resampling to the 0.6 \AA{} resolution of our Starburst99 models. Intervals in grey shaded regions were masked during the modeling process and include ISM emission and absorption features, chip gaps, and geocoronal Ly$\alpha$ and O{\sc~i}. The blue curve shows the median model (see text for details), while the teal dotted curves show 50 random samples from the cleaned MCMC chain. {\em Bottom:} Model residuals for the 50 samples.}
\label{fig:appendix}

\end{figure*}

\begin{figure*}[t]
\centering
\includegraphics[width=\linewidth,height=0.52\linewidth,trim={0.1cm 0.2cm 0.1cm 0.2cm},clip]{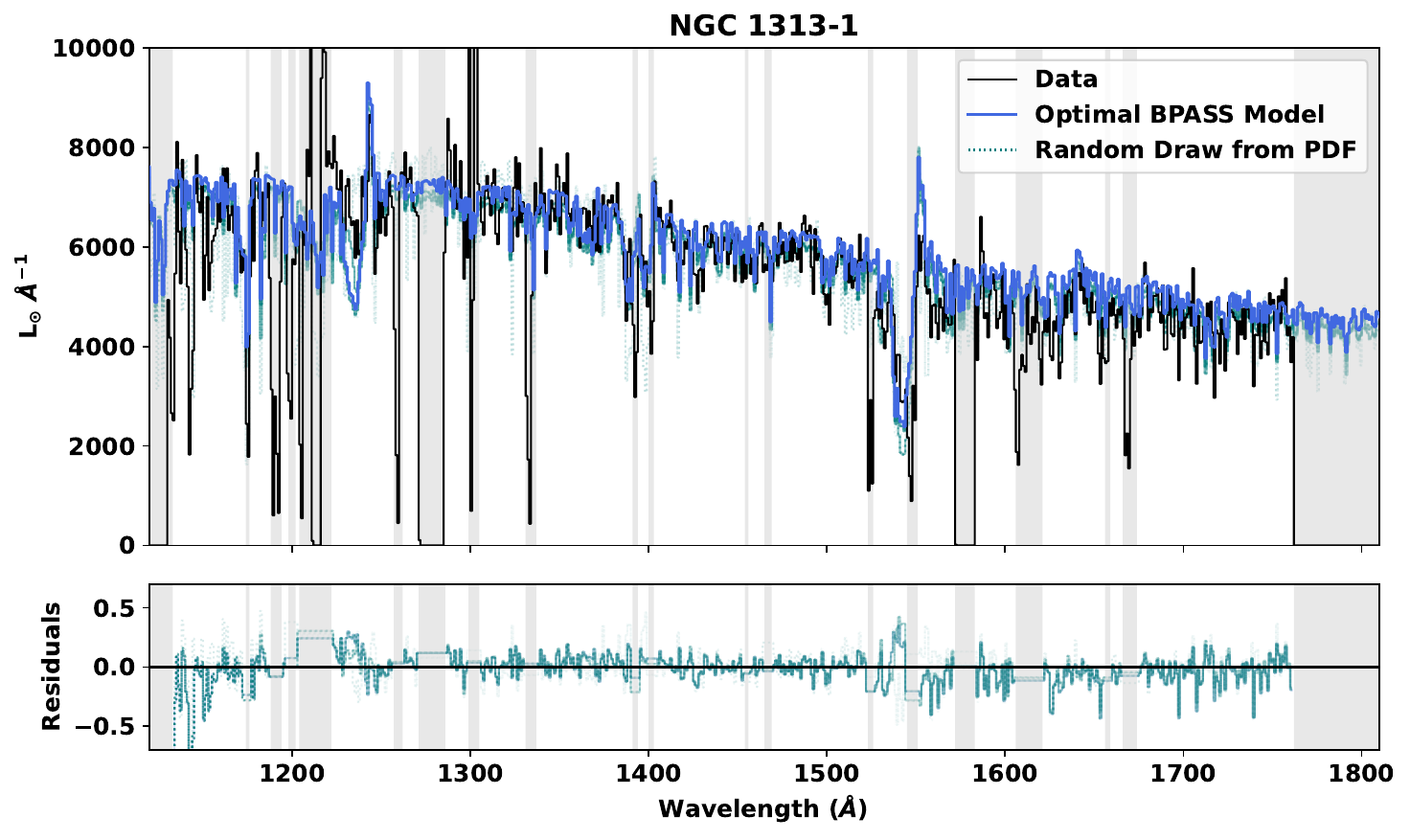}\\
\medskip
\includegraphics[width=\linewidth,height=0.52\linewidth,trim={0.1cm 0.2cm 0.1cm 0.7cm},clip]{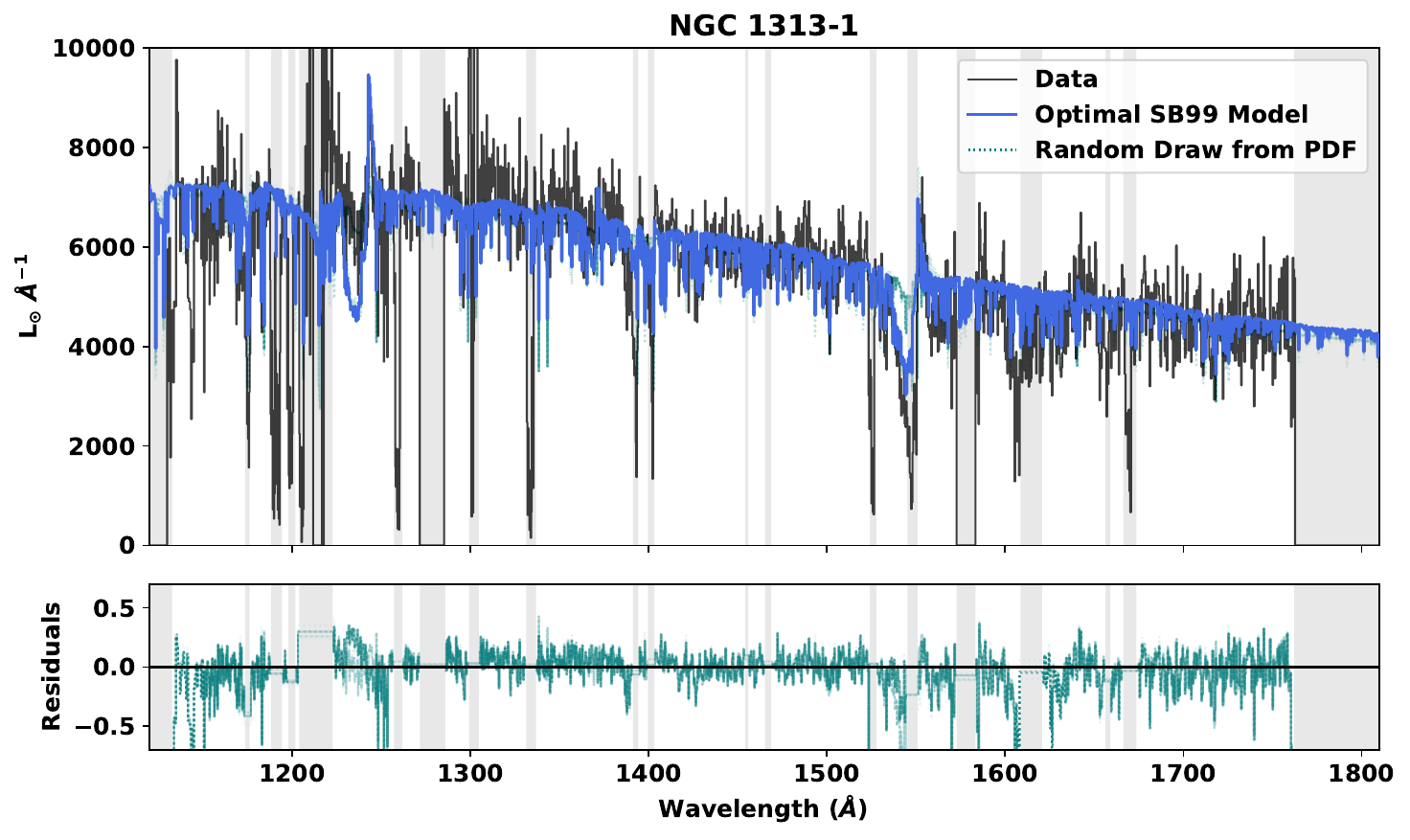}\\
\caption{Continuation of Fig. \ref{fig:appendix} for the subsolar metallicity YMC NGC 1313-1, as modeled using BPASS (top) and Starburst99 (bottom) SSPs.}

\end{figure*}

\begin{figure*}[t]
\centering
\includegraphics[width=\linewidth,height=0.52\linewidth,trim={0.1cm 0.2cm 0.1cm 0.2cm},clip]{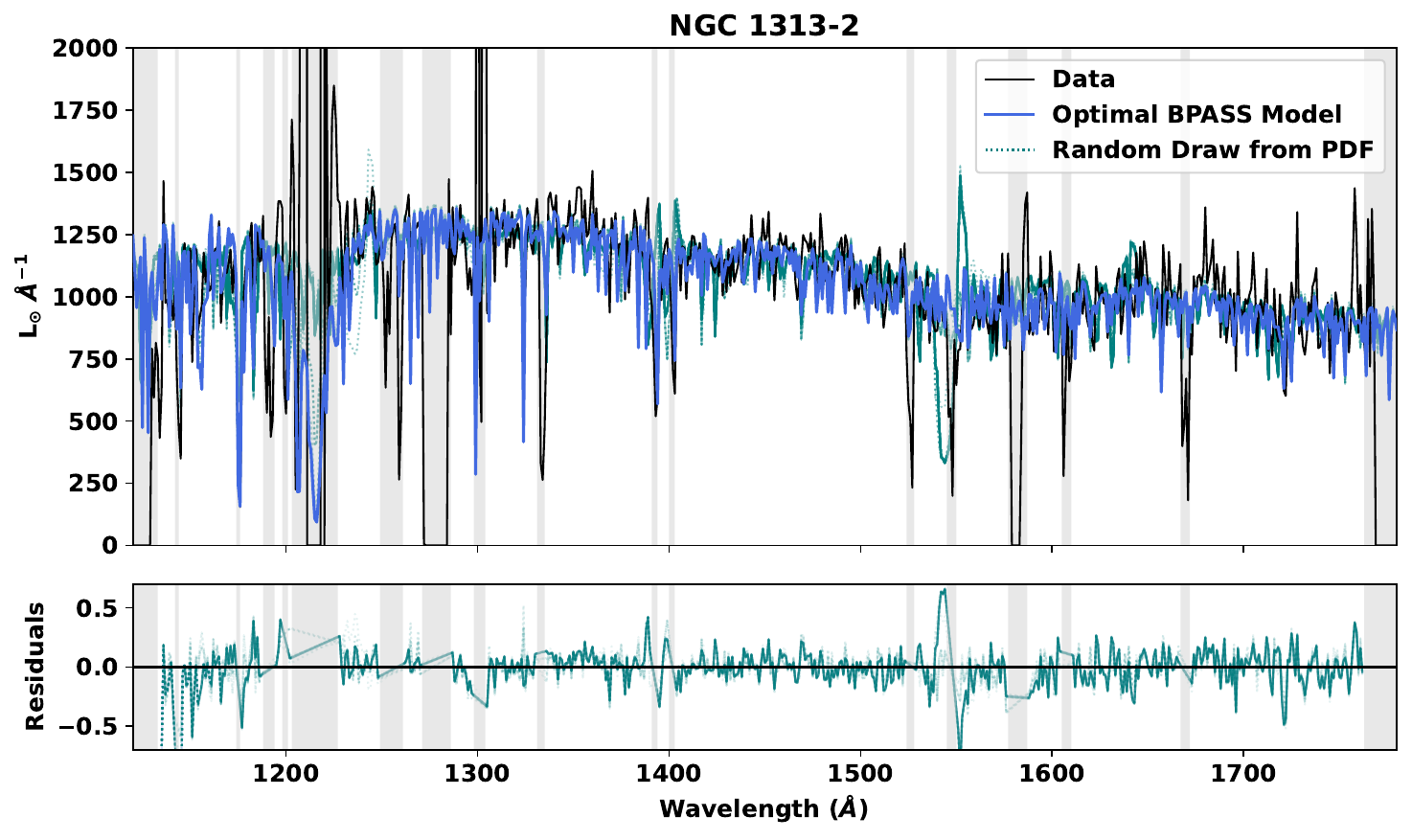}\\
\medskip
\includegraphics[width=\linewidth,height=0.52\linewidth,trim={0.1cm 0.2cm 0.1cm 0.7cm},clip]{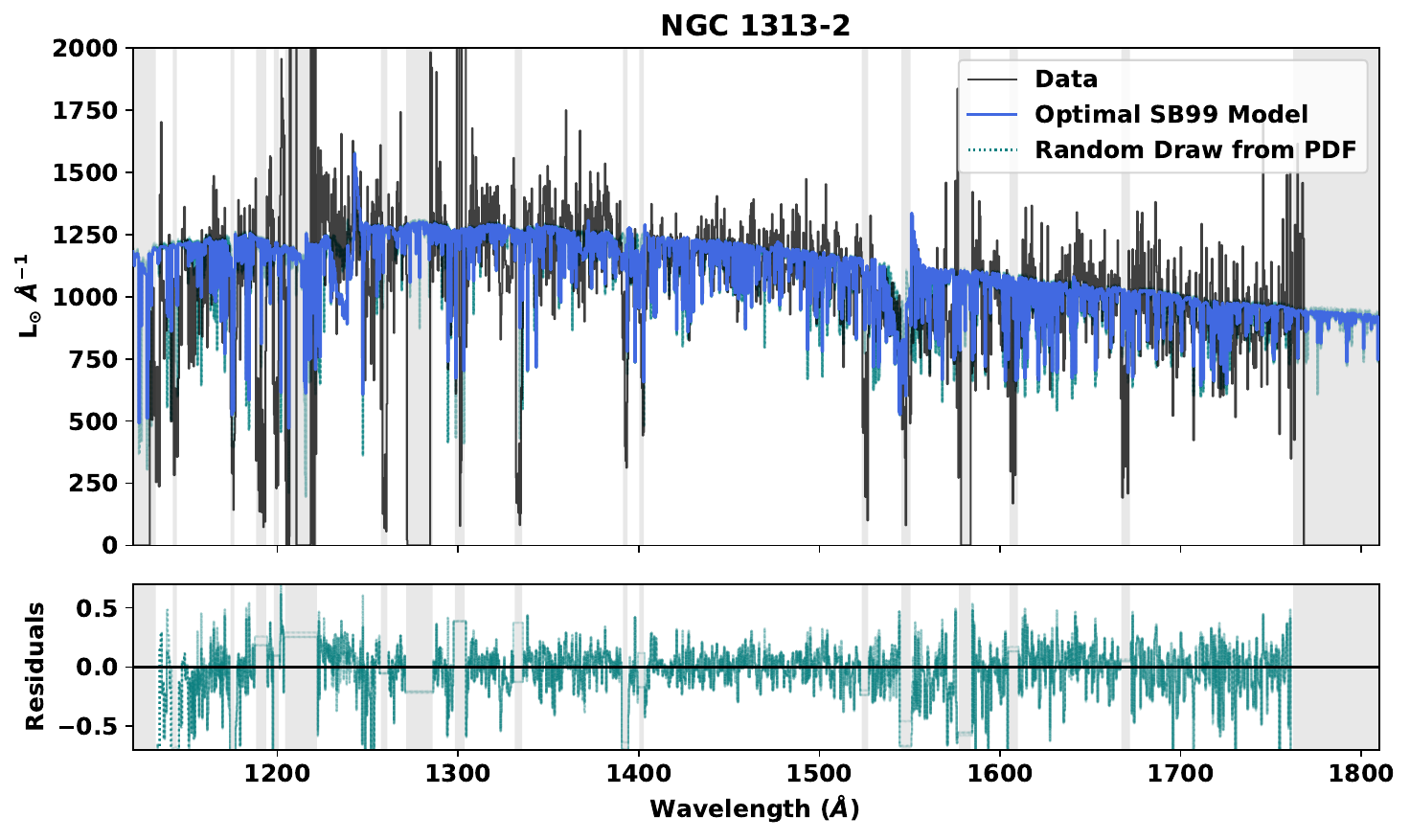}\\
\caption{Continuation of Fig. \ref{fig:appendix} for subsolar metallicity YMC NGC 1313-2, as modeled using BPASS (top) and Starburst99 (bottom) SSPs.}

\end{figure*}

\end{document}